\documentclass[]{aa}

\usepackage{natbib}
\usepackage{breqn}
\usepackage{xcolor}
\usepackage{graphicx}
\usepackage{txfonts}
\usepackage{multirow}
\usepackage{hyperref} 

\hypersetup{
    colorlinks = true,
    linkbordercolor = {cyan},
    linkcolor={blue},
    citecolor={blue},}

\begin{document}

\def\figureautorefname{Fig.}
\def\equationautorefname~#1\null{Eq. (#1)\null}

\def\sectionautorefname{Sect.}
\def\subsectionautorefname{Sect.}
\def\subsubsectionautorefname{Sect.}

\title{On the theoretical instability strips of $\gamma$-Doradus stars including the effect of the metallicity and rotation}
\titlerunning{Grid_Instabilities}

\author{L. Fellay\inst{1} \and M.-A. Dupret\inst{1} }
\institute{STAR Institute, University of Liège, 19C Allée du 6 Août, B$-$4000 Liège, Belgium}
\date{January,  2025}

\abstract
{Recent space missions such as \textit{Kepler} have provided large-scale observations of the  $\gamma$-Doradus instability strips (IS), which can be used to constrain models and explore their limitations. One persistent limitation is the prediction of the blue edge of the $\gamma$-Doradus IS, where a significant number of $\gamma$-Doradus stars are observed. Despite these observational advances, no systematic study has been undertaken to explore the effects of different physical processes on the $\gamma$-Doradus.}
{Our aim is to systematically explore the theoretical $\gamma$-Doradus IS with modern tools, accounting for the effects of rotation and metallicity,  providing a large grid of models and their oscillation parameters for the scientific community.}
{We investigated the non-adiabatic pulsation properties of stars in a grid of stellar models with masses between $1.35\,M_\odot$ and $2.5\,M_\odot$, metallicities between $Z=0.01$ and $Z=0.025$, and solid-body rotation rates ranging from $\Omega=0$ to $0.5\,\Omega_{\mathrm{crit}}$, where $\Omega_{\mathrm{crit}}$ denotes the critical rotation rate.}
{We find that, across all computations, the theoretical $\gamma$-Doradus IS agrees well with the observed IS, except in the extended blue region. In terms of radial orders, our models consistently reproduce the excited modes in broad agreement with observations for $\ell=1$, $\ell=2$, and Rossby modes. We also show that the range of excited radial orders is strongly dependent on the effective temperature.}
{The range of excited radial orders can be used to constrain the position of a star within the $\gamma$-Doradus IS and may provide insights into the physical mechanisms responsible for the discrepancies observed in the blue region. All computed grids are made available and include mode damping and growth rates for $\gamma$-Doradus stars, obtained with a time-dependent treatment of the convection–oscillation interaction.}

\maketitle

\noindent

\section{Introduction}
The instability strips (IS) of $\gamma$-Doradus stars were extensively investigated in the late 1990s and early 2000s to identify the driving mechanisms responsible for the observed pulsations and to reconcile theory with the observational constraints available at that time \citep{Kaye1999, Handler1999, Handler2002}. $\gamma$-Doradus stars, typically spanning spectral types A7 to F5, exhibit high-order $g$-mode instabilities driven by convective flux blocking at the base of their thin convective envelopes.  While the theoretical instability strips of $\gamma$-Doradus stars were studied in detail in the early 2000s, only a few works have revisited their theoretical description in light of the vastly improved observational constraints provided by past (\textit{Kepler}; \citep{Kepler2010}), current (\textit{TESS}; \citep{TESS2015}), and future (\textit{PLATO}; \citep{PLATO2014}) space missions. Most detailed non-adiabatic computations including time-dependent convection (TDC) have so far been based on relatively small grids of models \citep{Dupret2005,Grigahcene2005}, and later studies mainly focused on improved convection treatments and updated input physics \citep{Xiong2016}. On the observational side, space-based photometry has revealed large samples of $\gamma$~Dor and hybrid $\delta$~Scuti/$\gamma$-Doradus pulsators with rich gravity-mode spectra. In particular, \textit{TESS} has already delivered ensemble catalogues, extensive period-spacing samples, and detailed case studies of individual stars and clusters \citep{Antoci2019,Garcia2022a,VanReeth2022,Garcia2022b,Zhou2025a,Kliapets2025,Zhou2025b}. However, the effect of rotation on the $\gamma$-Doradus instability strip has not yet been explored systematically on large model grids. The influence of rotation has previously been addressed only for a limited number of models \citep{Bouabid2013} and without an exhaustive parameter study, and the excitation of Rossby ($r$) modes \citep{Saio2018} has not been investigated in the $\gamma$-Doradus context.  Regarding metallicity, previous work has considered its impact on the global IS boundaries without studying in detail the excitation mechanisms or making the corresponding model grids publicly available \citep{Grigahcene2006}.
\\~\\Accurate modelling of the instability strips of $\gamma$-Doradus stars (in particular the theoretical red edge) requires non-adiabatic stellar oscillation codes that includes a time-dependent treatment of convection–oscillation interactions \citep{Dupret2005,Xiong2016}. \citet{Guzik2000} demonstrated that modes in $\gamma$-Doradus stars can be driven by convective flux blocking at the base of their thin convective envelopes using the frozen-convection approximation.  However, subsequent work showed that only TDC models yield reliable physical interpretations of instability strips \citep{Dupret2005,Grigahcene2005}. Modern stellar oscillation codes such as \texttt{GYRE} \citep{GYRE2013, GYRE2018, GYRE2020} do not yet include a full TDC treatment, making it impossible to compute realistic instability strips or detailed mode damping/growth rates for $\gamma$-Doradus stars, since the frozen-convection approximation breaks down when the lifetime of convective eddies is shorter than the pulsation period.
\\~\\Comparisons between existing theoretical instability strips and observations show broad qualitative agreement, but several persistent discrepancies remain. Using the complete \textit{Kepler} sample of 611 $\gamma$-Doradus stars with detected period-spacing patterns, \citet{Li2019,Li2020} demonstrated that more oscillation modes are excited in observations than predicted by current theory, even when rotation is included. A further discrepancy,  is the presence of a sub-population of $\gamma$-Doradus stars located at the extreme blue edge or even outside the theoretical instability strip.   Similar tensions between theory and observations are emerging from large Kepler or TESS-based catalogues of $\gamma$-Doradus and hybrid pulsators, which reveal very rich and diverse mode spectra over a wide range of stellar parameters \citep{Antoci2019,Li2019,Li2020,Garcia2022a,Garcia2022b,Zhou2025a,Zhou2025b,Kliapets2025}. Ongoing and upcoming observations with \textit{TESS} and, in the longer term, \textit{PLATO} will provide additional constraints on these discrepancies and on the physical processes shaping the $\gamma$-Doradus instability strip.
\\~\\In this work and its associated materials, we present the first large-scale grid of stellar models for $\gamma$-Doradus stars specifically designed to investigate these discrepancies. The computations were performed with the non-adiabatic oscillation code \texttt{MAD} \citep{Dupret2001, Dupret2002} coupled to the stellar evolution code \texttt{CLES} \citep{Scuflaire2008a}. In addition,  to have a TDC treatment in \texttt{MAD}, we included the effects of rotation using the traditional approximation of rotation \citep{Bouabid2013}, which is well suited for gravito-inertial modes in stars without deep convective envelopes. This grid introduces several major improvements compared to pre-\textit{Kepler} works. First, the models adopt the AGSS09 solar abundance mixture \citep{Asplund2009}. Second, both \texttt{CLES} and \texttt{MAD} have been significantly updated over the past two decades, improving the reliability and precision of stellar structure and pulsation computations. Finally, modern computational resources allow large grids to be constructed, enabling a systematic exploration of physical effects (e.g. rotation, metallicity) on a global scale, rather than being limited to a few individual models as in earlier studies.
\\~\\In our grid, we considered stellar masses between $1.35\,M_\odot$ and $2.5\,M_\odot$, metallicities from $Z=0.01$ to $Z=0.025$, and initial rotation rates from $\Omega=0$ to $0.5\,\Omega_{\mathrm{crit}}$, where $\Omega_{\mathrm{crit}}$ is the critical rotation rate. Rotation was included only in the oscillation computations, and not in the stellar evolution models. In these computations, rotation was assumed to be solid-body rotation. We chose not to include rotational effects during the evolution, as their treatment remains uncertain and would introduce dependencies on arbitrary modelling choices. Instead, we provide grids computed for different rotation rates, allowing the user to select the desired rotation rate for the oscillation calculations, for instance based on evolutionary models. For each model, we provide the range of radial orders with unstable modes, the corresponding mode frequencies/periods and damping/growth rates, as well as detailed stellar structure properties. This grid is intended as a resource for the observational community, enabling more detailed studies of stellar interiors and populations across the $\gamma$-Doradus IS. It also provides a modern reference for the asteroseismic community, particularly regarding the theoretical instability strip of $\gamma$-Doradus stars and the damping/growth rates of their modes.
\\~\\This article is organised as follows. In \autoref{sect_gammaDor} we explore the impact of different stellar model parameters on the $\gamma$-Doradus IS. In \autoref{sect_impact_ortation_gammaDor} we investigate the role of rotation.  Finally, in \autoref{sect_discussion_conclusion} we discuss our results and present our conclusions.

\section{$\gamma$-Doradus stars and convective flux blocking}\label{sect_gammaDor}
\begin{figure}[h]
\centering
\includegraphics[width=\hsize]{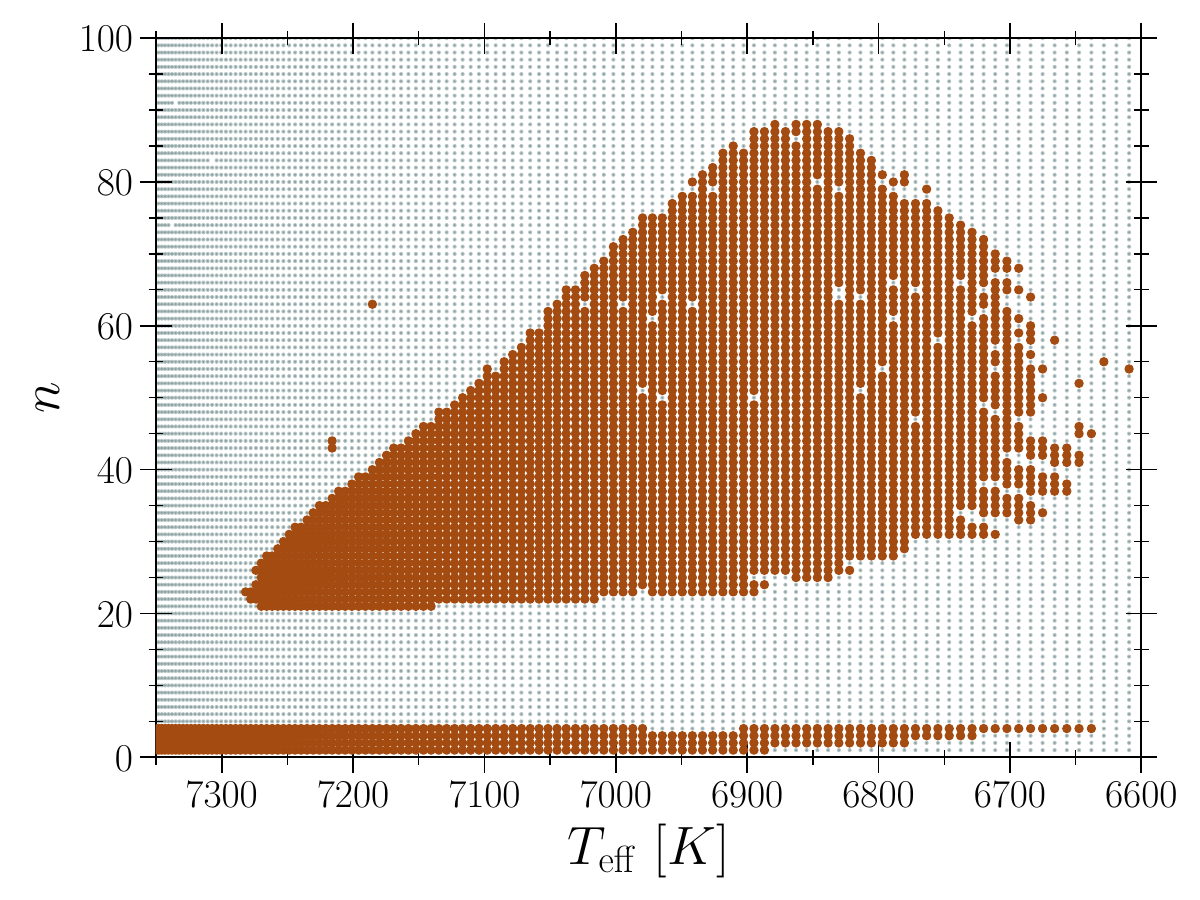}
\caption{Radial orders of the stable and unstable $\ell=1,\,m=0$ modes for a $1.55\ M_\odot$ track. Each grey dot corresponds to a stable mode, while orange dots denote unstable modes. The detail description of the physical parameters used in the stellar models is given in \autoref{sect_impact_gammaDor}.}
\label{fig._IS_n_Teff_reference}
\end{figure}

In the case of $\gamma$-Doradus stars, the driving mechanism responsible for the observed $g$-mode instabilities is the convective flux blocking that operates at the base of their thin, near-surface convective envelopes \citep{Guzik2000,Dupret2005,Grigahcene2005,Xiong2016}. Physically, this mechanism arises because the convective envelope cannot adjust instantaneously to the periodic perturbations induced by the pulsations. As a result, part of the radiative flux that would normally be transported outward by convection is temporarily trapped at the base of the convective zone at the hot phase of the cycle.   During the pulsation cycle, this trapped thermal energy is periodically converted into work,  providing net driving as part of a thermodynamic heat engine. The classical picture of convective flux blocking, following \citet{Dupret2005},  is that the efficiency of this mechanism,  and thus the excitation of unstable modes,  is mainly controlled by the location of the base of the convective envelope relative to the so-called transition region, where the pulsation period is of the same order as the local thermal relaxation timescale. When the base of the convective zone lies within this transition region, the thermodynamic heat engine has a period close to that of the g-modes,  the thermodynamic heat engine is the most efficient, leading to the driving.  If, on the other hand, the base of the convective zone is situated well above or below the transition region, radiative damping dominates and the modes remain stable.
\\~\\In this section, we explore how the balance between driving and damping determines the range of radial orders and periods of the excited oscillation modes.  For a $1.55\,M_\odot$ evolutionary track (see \autoref{sect_impact_gammaDor} for a detailed description of the physical parameters used in the stellar models), we illustrate in \autoref{fig._IS_n_Teff_reference} the range of radial orders corresponding to stable and unstable modes as a function of the effective temperature.
\\~\\This figure shows the range of radial orders at which unstable modes are found in classical $\gamma$-Doradus stars. In general, unstable gravity modes are found with radial orders between $n \simeq 20$ and $n \simeq 90$, with lower radial orders ($n \sim 20$) favoured near the blue edge of the instability strip and higher radial orders favoured near the red edge. This theoretical range can be compared with the observed radial orders reported by \citet{Li2019}. Overall, the agreement between theory and observations is satisfactory, although the observed instability strip appears more extended in radial order, with a non-negligible number of observed $\ell=1$ modes at $n<20$ and up to $n \sim 100$.  Low-radial-order unstable gravity modes ($n<10$) are also visible in this diagram; these correspond to the $\delta$~Scuti instability strip, which will be discussed in a separate article. 
Compared to the work of \citet{Bouabid2013}, we find a larger number of excited modes in the central region of the instability strip. The models explored by \citet{Bouabid2013} were likely concentrated towards the blue edge, which naturally leads to the excitation of predominantly low-radial-order modes.

\subsection{On the low radial order/period limit}
\begin{figure}[h]
\centering
\includegraphics[width=\hsize]{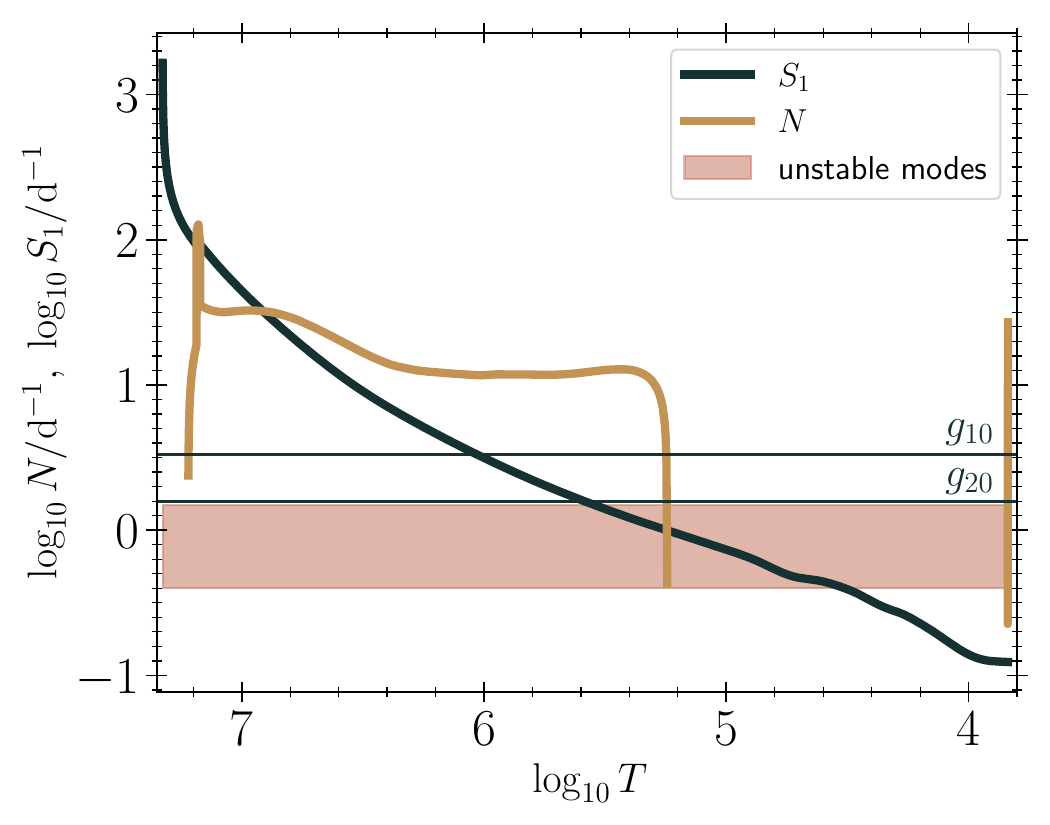}
\caption{Propagation diagram for a $1.55 M_\odot$ $\gamma$-Doradus star with $T_{\rm eff}=7000 {\rm K}$. The yellow and black lines corresponds to respectively the Brunt--V{\"a}is{\"a}l{\"a}  frequency $N$ and the Lamb frequency $S_\ell$ of the $\ell=1$ modes. The orange region corresponds to the frequency at which unstable modes are found in this model (with radial orders between $21$ and $76$), the black horizontal lines corresponds to the frequency of the $n= 10$ and $n=20$ g-modes which are stable.}
\label{fig.illustration_convective_blocking}
\end{figure}

First, a region of stable modes exists between the $\delta$-Scuti unstable region (low radial orders) and the $\gamma$-Doradus unstable region (higher radial orders). Two arguments can account for this stabilisation region. The first concerns the evanescent zone. For a $g$~mode to propagate, its angular frequency must be lower than both the Brunt-V{\"a}is{\"a}l{\"a} frequency and the Lamb frequency.  Mathematically,  the squared Lamb frequency $S_\ell$ is defined as
\begin{equation}\label{P31_eq_Lamb}
S_\ell^2 = \frac{\ell(\ell+1)c^2}{r^2},
\end{equation}
where $c$ is the local sound speed and $r$ is the local radius.  The squared Brunt-V{\"a}is{\"a}l{\"a} frequency is defined as
\begin{equation}\label{P31_eq_BV}
N^2 = g\left(\frac{1}{\Gamma_1}\frac{d\ln P}{dr} - \frac{d\ln \rho}{dr} \right).
\end{equation}
where $\Gamma_1=\left(\frac{\partial\ln P}{\partial \ln \rho}\right)_S$ is the first adiabatic exponent,  $g$ is the gravitational acceleration,  $P$ is the pressure and $\rho$ the density.  $N^2$ is positive in the radiative zones and negative in the convective zones. When inspecting the propagation diagram of a typical $\gamma$-Doradus model (see \autoref{fig.illustration_convective_blocking}), we find that low-radial-order $g$~modes must traverse a wide evanescent region before reaching the potential excitation layer at the base of the convective envelope. In this case, the modes are strongly attenuated while travelling through the evanescent zone, so that radiative damping dominates and they remain stable. By contrast, higher-radial-order $g$~modes have propagation cavities that are better connected to the excitation region and can therefore overcome the damping and become unstable \citep{Dupret2005,Xiong2016}.
The second explanation arises from the behaviour of the eigenfunctions themselves. Near the surface, the Lagrangian pressure perturbation $\delta P$, following \citet{Buta1979}, can be approximated by:
\begin{equation} \label{eq_approx_amplitude}
\frac{\delta P}{P}= \left[\ell(\ell+1) K - 4 -K^{-1} \right]\frac{\xi_r}{r},
\end{equation}
where $K$ is the square of the inverse dimensionless frequency of the mode:
\begin{equation}
K=\frac{GM}{\sigma^2 R^3}.
\end{equation}
At the surface,  the range of dimensionless frequencies corresponding to the intermediate stable region coincides with the point where the right-hand side of \autoref{eq_approx_amplitude} approaches zero. As a result, the amplitude of the modes near the surface remains small and cannot be efficiently excited by convective blocking. Both explanations are plausible, and it is not currently possible to distinguish between them; in both cases, the affected modes remain stable.

\subsection{On the high radial order/period limit}
\begin{figure}[h]
\centering
\includegraphics[width=\hsize]{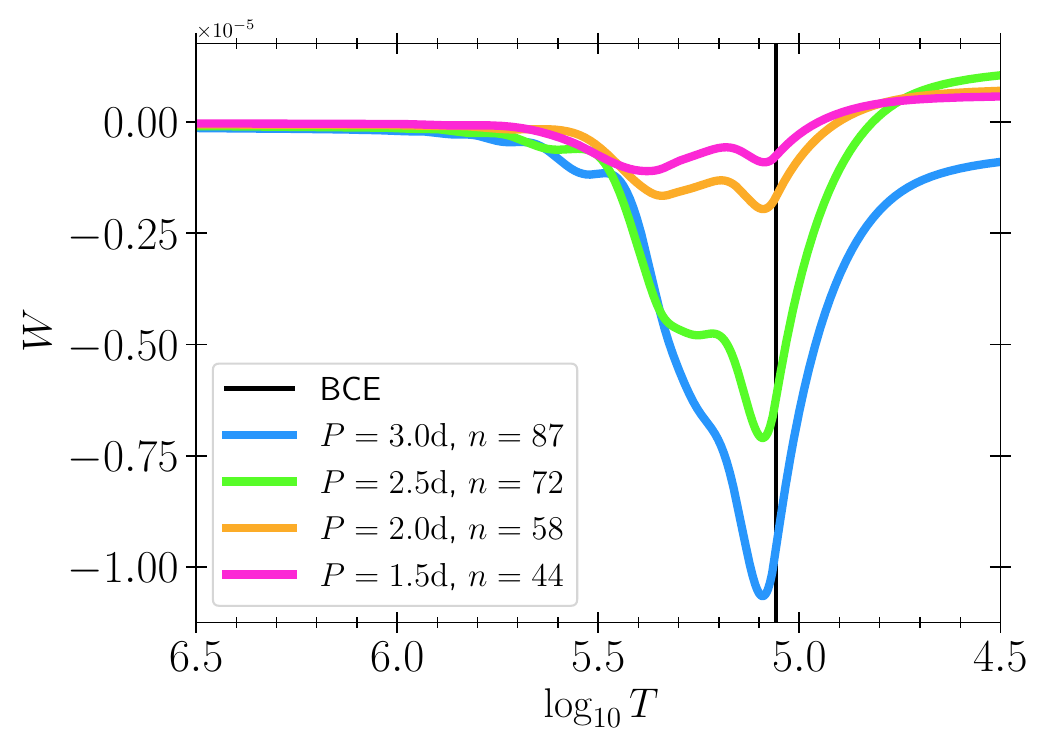}
\caption{Work integral for selected stable and unstable $\gamma$-Doradus g modes as a function of $\log_{10} T$ within a $1.55,M_\odot$ model with $T_{\rm eff}=7000 {\rm K}$. The black vertical line marks the location of the base of the convective envelope (BCE). Each colour corresponds to a different mode (i.e. a different period). The net damping or driving is given by the surface value of the work integral: positive values indicate unstable (excited) modes, while negative values correspond to damped modes. A decreasing work integral indicates damping, whereas an increasing work integral indicates driving.}
\label{fig.illustration_blocking_W}
\end{figure}

\begin{figure}[h]
\centering
\includegraphics[width=\hsize]{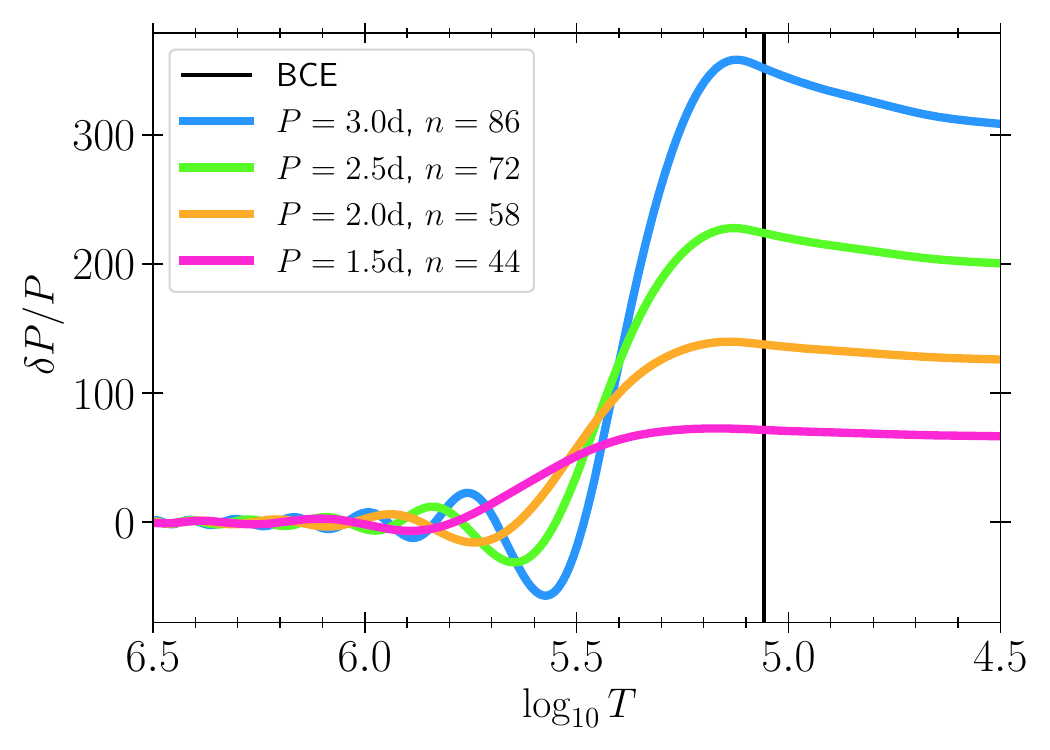}
\caption{$\delta P/P$ eigenfunctions for the same modes presented in \autoref{fig.illustration_blocking_W} as a function of the temperature of the different layers of a $1.55 M_\odot$ model with $T_{\rm eff}=7000K$. Both the near core region and the surface were cut in this figure to focus on the region close to the BCE where the driving and damping of the oscillations takes place.}
\label{fig.illustration_blocking_deltaPP}
\end{figure}

\begin{figure}[h]
\centering
\includegraphics[width=\hsize]{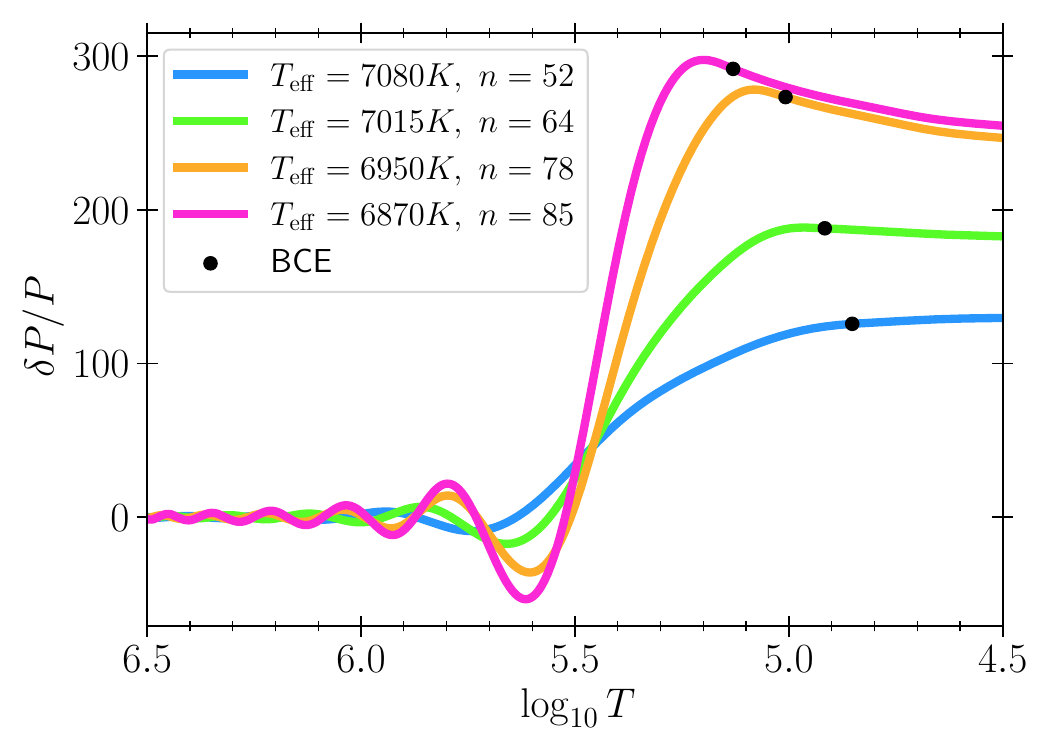}
\caption{$\delta P/P$ eigenfunctions of the unstable mode with the highest period across a few models in the $1.55 M_\odot$ IS. Both the near core region and the surface were cut in this figure to focus on the region close to the BCE where the driving and damping of the oscillations takes place.}
\label{fig.illustration_blocking_deltaPP_Teff}
\end{figure}

\begin{figure}[h]
\centering
\includegraphics[width=\hsize]{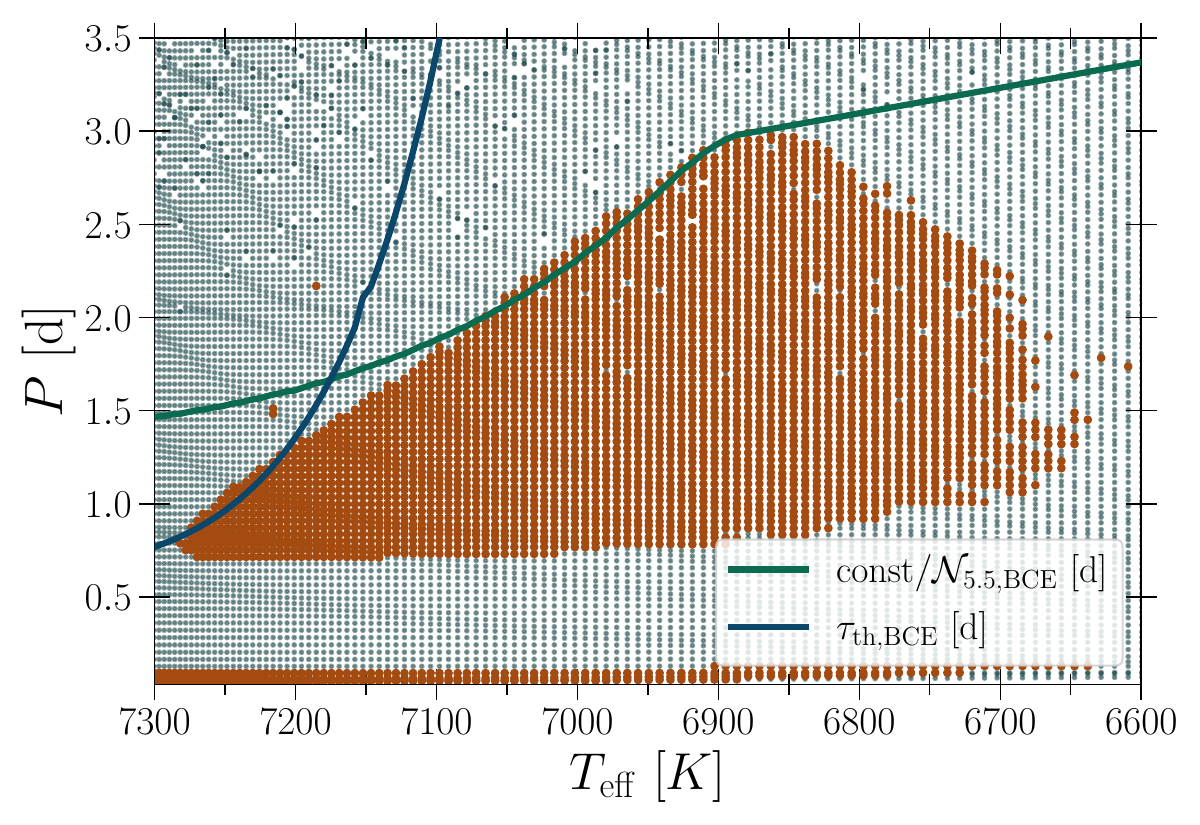}
\caption{Periods of the stable and unstable $\ell=1,\,m=0$ modes for the reference $1.55\,M_\odot$ track. Each black dot corresponds to a stable mode, while red dots denote unstable modes. $\tau_{\mathrm{th., BCE}}$ is the thermal relaxation timescale at the BCE. For this sequence, the constant value of $\mathcal{N}_{5.5,\mathrm{BCE}}$ was determine to be equal to two.}
\label{fig._IS_P_Teff_reference}
\end{figure}

A second important aspect of mode excitation concerns the range of radial orders and periods that are effectively driven by convective luminosity blocking. As mentioned previously, this range depends strongly on the stellar effective temperature. This dependence is particularly striking near the blue edge of the instability strip, where only a few tens of modes ($\sim 30$) are excited, compared with $60$--$70$ modes in the middle of the IS. The detailed explanation for these ranges is far from trivial, because several effects shape the final stability of the modes: changes in the efficiency of convective blocking, the strength of radiative damping, and the impact of stellar evolution on the internal structure and on the excitation/damping mechanisms.
\\~\\Several possibilities were considered to explain this behaviour, and we find that the depth of the base of the convective envelope (BCE) largely controls the excitation efficiency. First, it sets the region where convective blocking becomes possible: the blue edge of the instability strip is systematically defined by models in which the BCE lies within the transition zone, where the local thermal timescale at the BCE $\tau_{\mathrm{th., BCE}}$ is comparable to the mode period. However, this location does not correspond to the region where convective blocking is most efficient (more than $75\,\%$ of the IS lies outside this regime), but rather to the point where it becomes sufficiently efficient to overcome radiative damping for low–radial-order modes. Following the results of \citet{Godart2009}, radiative damping scales approximately as $\sigma^{-4}$ and therefore becomes increasingly efficient for higher–radial-order modes. As a consequence, low–radial-order modes are more easily driven when convective blocking is still relatively weak. To illustrate this phenomenon, \autoref{fig.illustration_blocking_W} shows the work integrals of modes with different periods (and radial orders) for a $1.55\,M_{\odot}$ model with $T_{\rm eff}=7000~\mathrm{K}$.
\\~\\As shown in this figure, several effects contribute to the driving and damping of the modes. As expected, the radiative damping strengthens with increasing period, as indicated by the decreasing work integral, which stabilises long-period modes. Most work integrals also exhibit a non-negligible driving contribution around $\log_{10}T=5.2$, which was initially unexpected and corresponds to the $\kappa$-mechanism associated with the iron-group opacity bump \citep{Cox1966,Pamyatnykh1999}. This contribution is stronger at shorter periods, again favouring the excitation of low–radial-order modes. The dominant driving, however, arises close to the BCE, and the efficiency of this convective blocking contribution clearly depends on the mode period. To clarify this dependence and the enhanced efficiency of the $\kappa$-mechanism at low periods, \autoref{fig.illustration_blocking_deltaPP} displays the eigenfunctions of the same modes.
\\~\\A comparison of the eigenfunctions shows that increasing the mode period shifts the last radial node of the eigenfunction closer to the BCE. As a result, the location of the maximum amplitude or the plateau in $\delta P/P$ varies with the period. In particular, for low-frequency modes the amplitude maximum occurs far below the $\log_{10}T=5.2$ region where the $\kappa$-mechanism acts, whereas for higher-period modes the maximum lies close to or beyond this region. The efficiency of the $\kappa$-mechanism is therefore strongly linked to the behaviour of the eigenfunctions near the opacity bump. Since the $\kappa$-mechanism contributes to the driving of the modes, variations in opacity and metallicity will influence both the number of excited modes and their periods; this impact is investigated in the next section. In addition, shifting the last node closer to the BCE with increasing period also affects the efficiency of convective blocking, as the behaviour of the eigenfunctions near the BCE plays a crucial role in the driving. To confirm this conjecture, we examined the eigenfunctions of the highest-period unstable modes across several models along the $1.55,M_\odot$ evolutionary track, shown in \autoref{fig.illustration_blocking_deltaPP_Teff}.
Despite clear differences between the stellar models and the radial orders of their highest-period unstable modes, the eigenfunctions display remarkably similar behaviour below the BCE.  In all cases, the minimum preceding the last node lies near $\log_{10}T \sim 5.6$, and the maximum amplitude lies just below $\log_{10}T \sim 5.2$, immediately before the BCE. Since the eigenfunctions exhibit similar behaviour close to the BCE, the phase accumulated by the modes in this region should also be similar. We therefore define the local dimensionless buoyancy radius between the layer at $\log_{10}T=5.5$ and the BCE as
\begin{equation}
\mathcal{N}_{5.5,\mathrm{BCE}}
\equiv
\frac{1}{\sigma}
\int_{r_{\log_{10}T=5.5}}^{r_{\mathrm{BCE}}}
\frac{N(r')}{r'},\mathrm{d}r',
\end{equation}
where $\sigma$ is the frequency of the longest-period unstable mode. This quantity measures the contribution of the stellar structure in this region to the phase accumulated by the mode. The similar behaviour of the eigenfunctions near the BCE therefore implies that
\begin{equation}
\mathcal{N}_{5.5,\mathrm{BCE}}
\simeq \mathrm{const}.
\end{equation}
The value of this constant is not known a priori, but it is approximately the same for all models belonging to a given evolutionary sequence. In practice, we determine it by evaluating the above expression for one reference model. This relation expresses the fact that the local behaviour of the longest-period unstable modes is approximately equivalent in the region near the BCE. This interpretation is supported by the numerical results presented in \autoref{fig.illustration_blocking_deltaPP_Teff}.
This evolutionary trend is visible in \autoref{fig._IS_P_Teff_reference}, which shows the periods of stable and unstable modes as a function of effective temperature for the $1.55,M\odot$ sequence.
In this figure we compare the modelled unstable periods with the various theoretical limits derived analytically \citep{Dupret2005,Dupret2009,Xiong2016}. The blue edge of the instability strip is clearly associated with the region where the thermal relaxation timescale at the BCE matches the mode period. As the star evolves, the period of the longest unstable modes increases, and the slope of this transition is governed by the evolution of the buoyancy radius in the vicinity of the BCE, as discussed previously.  Finally, a second transition occurs, in which the maximum unstable period decreases significantly. This happens when the BCE migrates clearly outside the transition region and the efficiency of convective blocking declines which happen for stars with $T_{\rm eff}<6900,{\rm K}$.  This regime is difficult to characterise precisely, because it depends sensitively on the adopted time-dependent convection–oscillation interaction formalism \citep{Dupret2005,Houdek2015}.  From an observational point of view, the longest observed unstable periods can place constraints on the buoyancy radius in the BCE region for stars located in the blue part of the instability strip \citep{VanReeth2015,Li2019}. Finally, the wavy pattern in \autoref{fig._IS_P_Teff_reference} corresponds to dips in the period spacing produced by mode trapping in the vicinity of the convective core. These glitches are observed in $\gamma$-Doradus stars and provide valuable constraints on the structure of the near-core region.

\section{Impact of different parameters on the theoretical $\gamma-$Dor IS }\label{sect_impact_gammaDor}
\begin{figure}[h]
\centering
\includegraphics[width=\hsize]{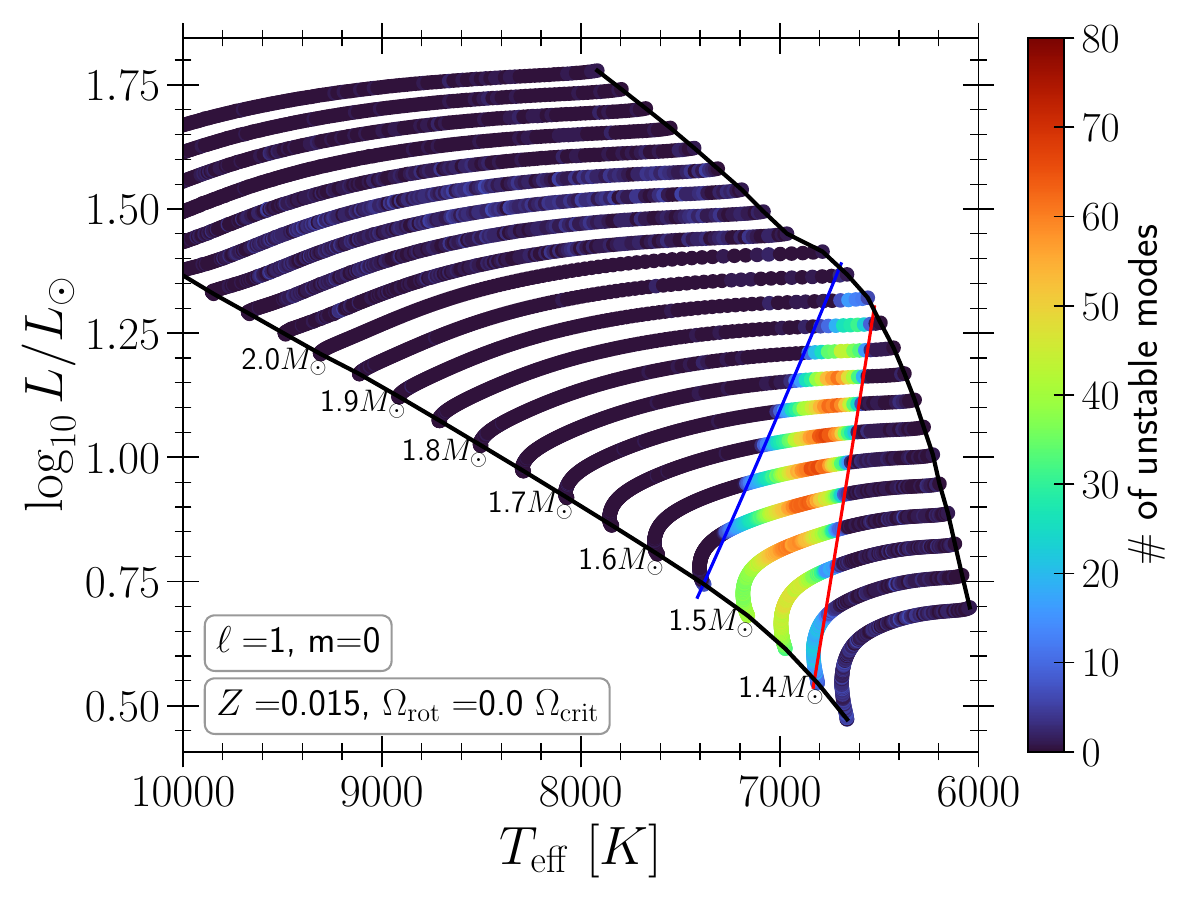}
\caption{HR-diagram of the $\gamma-$Doradus IS for our reference grid. The color code corresponds to the number of unstable $\ell=1$ $g$ modes for each model in each evolutionary sequence. For comparison, on the observational side, $\gamma$-Doradus stars are found at effective temperatures up to $9500 \mathrm{K}$, although the highest density of observed $\gamma$-Doradus stars remains close to the IS shown in this figure \citep{Li2019}.}
\label{fig.reference_IS_gammaDor}
\end{figure}

Using the classical view of the problem, the position of the instability strip should be particularly dependent on the mixing-length parameter chosen in the stellar models and all the parameters related to the convection. \citet{Dupret2005} pointed out that, to reproduce the observations available at the time, the best-suited mixing-length theory \citep[MLT,][]{Cox1968}  convective parameter $\alpha_{\mathrm{MLT}}$ was $\alpha_{\mathrm{MLT}} = 2.0$ in CLES. At that time, the standard chemical composition used for all such computations in the literature was a metallicity of $Z = 0.02$, adopting the GN93 solar abundances \citep{GN1993}. Even though metallicity is not the dominant factor affecting the instability strip, it directly affect stellar evolution and the efficiency of the $\kappa-$mechanism responsible for part of the driving.  In addition, rotation also impacts the instability strips of these stars, stabilising modes with particular azimuthal orders $m$ and shifting their frequencies in the observer frame, the effect of the rotation is discussed in \autoref{sect_impact_ortation_gammaDor}. 
\\~\\In this section we are exploring the effect of different modifications of the stellar structure and physical ingredients starting from a 'reference' set of stellar ingredients kept constant for the computations of a reference instability trip across the HR diagram. Unless precised in the main text, the physical ingredients used to compute the instability strip remain the same as our reference grid.  All the evolutionary sequences were computed with the Code Liégeois d'Evolution Stellaire \citep[CLES,][]{Scuflaire2008a}. For the reference grid of models we used an initial hydrogen mass fraction of $X=0.72$, a solar metallicity $Z=0.015$, a MLT convective parameter of $\alpha=2.0$, a step-like overshooting of $\alpha_{\mathrm{over}}=0.2$ (only acting in the core convective zone).  The models were stopped when the central hydrogen mass fraction reached $X_C=0.05$.  For all the stellar modelling we used the AGSS09 abundances \citep{Asplund2009}, the FreeEOS equation of state \citep{Irwin2012}, the OPAL opacities \citep{Iglesias1996}, the $T(\tau)$ relation from Model-C of \citet{Atmosphere1981} for the atmosphere, and the nuclear reaction rates of \cite{Reaction2011}.  
\\~\\For the reference grid of models,  \autoref{fig.reference_IS_gammaDor} illustrates the associated HR diagram of the $\gamma-$Doradus IS as computed with MAD for $\ell=1$ gravity modes.
For this reference grid, the $\gamma-$Doradus IS is located for models with an effective temperate  between $7300 K$ and $6600K$ and luminosities $log_{10}(L/L_\odot)$ between $0.6$ and $1.3$. When comparing this IS to the one obtained by \cite{Dupret2005},  the results seems compatible and the IS is roughly in the same position on the HR-diagram, which was expected as we did not modified any treatment of the convection/oscillations interactions.  Equations defining the theoretical blue edge and red edge of this reference grid $\gamma-$Doradus IS are provided in \autoref{apx_eq_edges_IS}.

\subsection{Impact of the metallicity}\label{subsubsect_metallicitygamma}
\begin{figure}[h]
\centering
\includegraphics[width=\hsize]{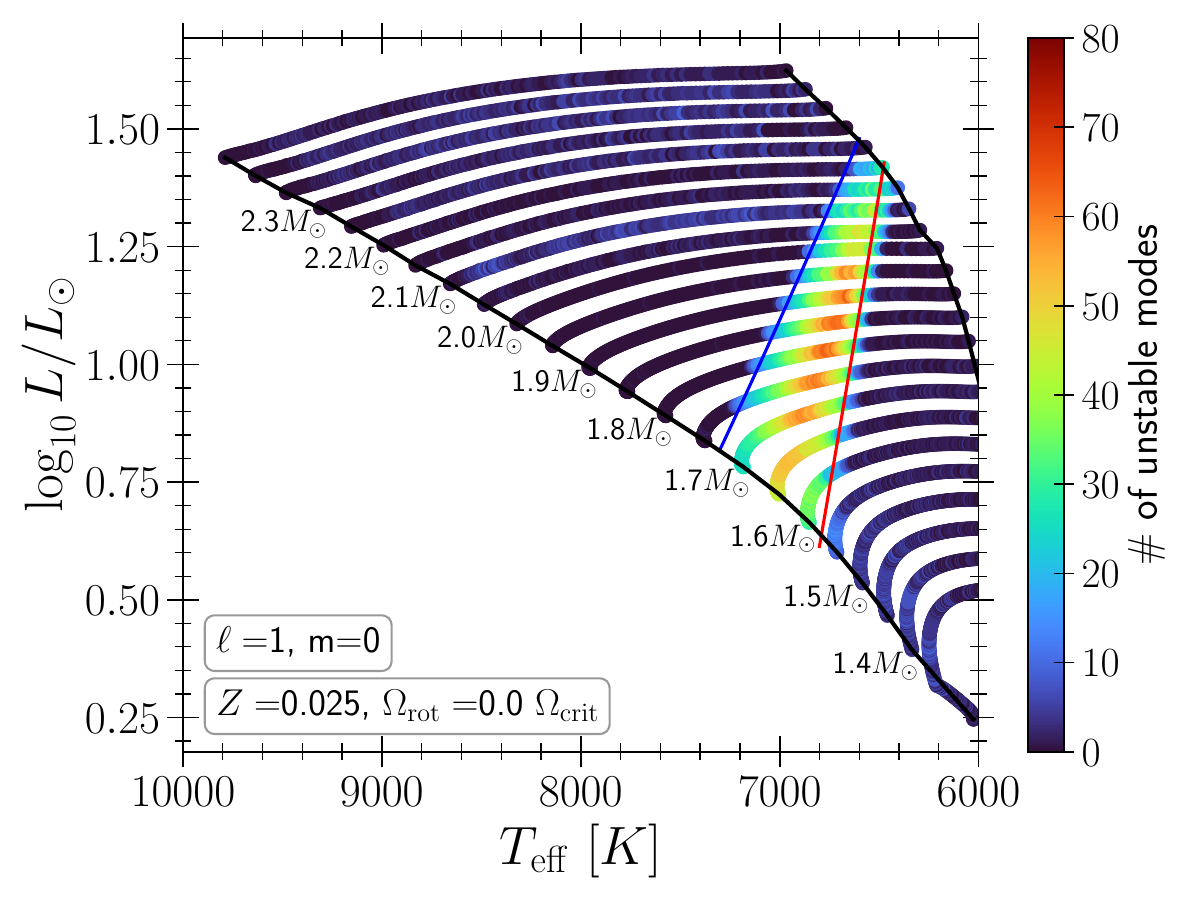}
\caption{HR diagram of the $\gamma$-Doradus IS for a grid with metallicity $Z=0.025$. The colour scale indicates the number of unstable $\ell=1$ $g$~modes for each model in each evolutionary sequence. }
\label{fig.gammaDor_Z}
\end{figure}

\begin{figure}[h]
\centering
\includegraphics[width=\hsize]{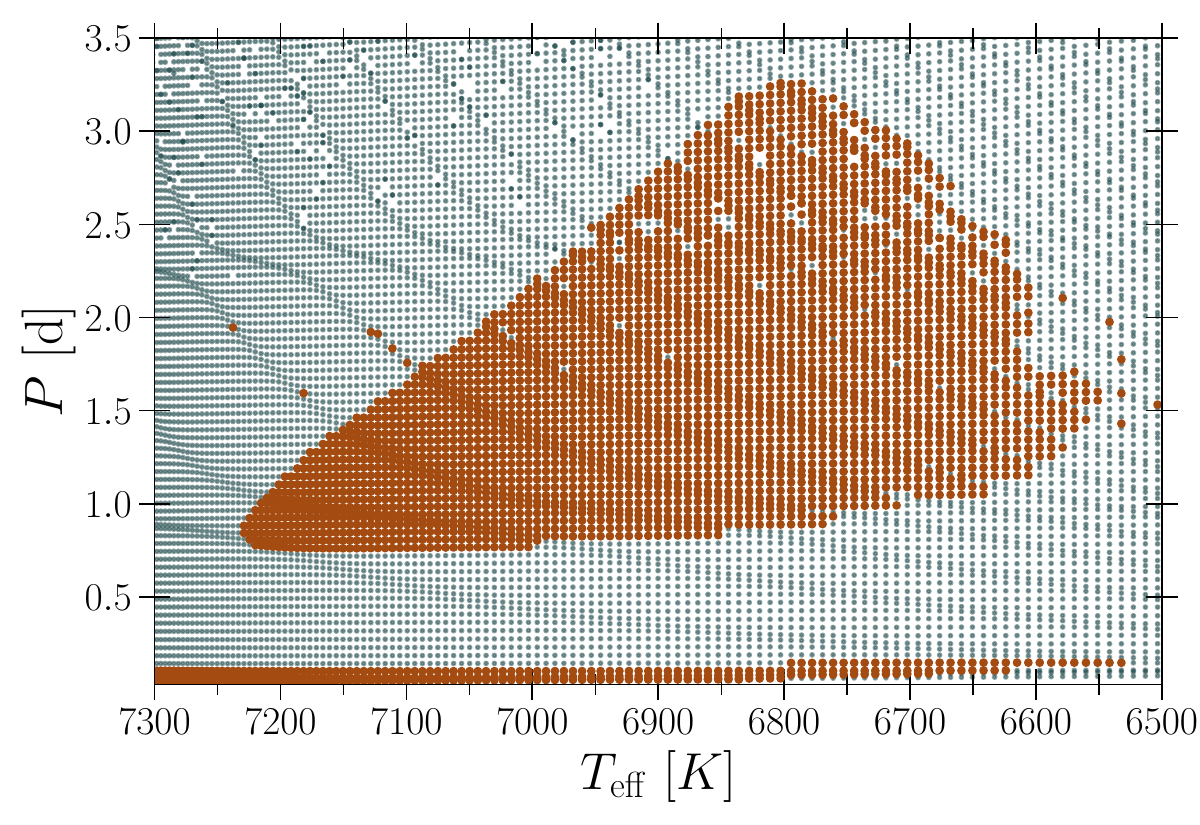}
\caption{Periods of the stable and unstable $\ell=1,\,m=0$ modes for the $1.75\,M_\odot$ track with $Z=0.025$.  Each black dot corresponds to a stable mode,  while red dots denote unstable modes.}
\label{fig._IS_P_Teff_Z0.025}
\end{figure}

To explore the instability strip of $\gamma$-Doradus stars in more detail, we computed a grid of models with masses ranging from $1.35\,M_\odot$ to $2.5\,M_\odot$ in steps of $0.05\,M_\odot$, and metallicities from $Z=0.010$ to $Z=0.025$ in steps of $0.005$. In \autoref{fig.gammaDor_Z}, we illustrate the $\gamma$-Doradus IS for a metallicity of $Z=0.025$, which can be compared to the reference case at $Z=0.015$.
\autoref{fig.gammaDor_Z} shows that, as expected \citep{Grigahcene2006}, the position of the instability strip in the HR diagram is not strongly affected by metallicity. For comparison, the red-edge and blue-edge used in this figure are the same as for the reference grid. We only see a small increase of the size of the IS of about $50K$ in each direction, which is expected as the amount of metals increases and so is the strenght of the $\kappa$ mechanism.  In terms of stellar evolutionary impact models with higher metallicity reach the instability strip at higher masses and younger ages compared to low-metallicity models.  Similarly, decreasing metallicity shifts the instability strip to lower stellar masses. Exploring cases of extremely low metallicities, such as those that could potentially in globular clusters, could therefore be of particular interest, as $\gamma$-Doradus stars in these environments may correspond to low-mass stars still on the main sequence despite the advanced age of the clusters.
\\~\\In addition to the effect of the metallicity on the stellar structure,  the driving of the modes should also be affected as the efficiency of the $\kappa-$mechanism should be impacted by the metallicity. On a global scale we do not see a particularly significant impact on the amount of unstable modes found. However, their period range can be modified, in \autoref{fig._IS_P_Teff_Z0.025} we illustrate the range of period at which unstable modes are found in a $1.75\,M_\odot$ track with $Z=0.025$.
By comparing the period range of unstable modes in \autoref{fig._IS_P_Teff_Z0.025} and for the solar metallicity case in \autoref{fig._IS_P_Teff_reference} we see an effect of the metallicity.  The modes driven in the high metallicity case are shifted toward higher periods, which is logical as the efficiency of the $\kappa-$mechanism is increased favouring the driving of higher period modes. We also note that the stellar evolution has a similar effect on the modes periods, it is therefore difficult to distinguish the two effects as metallicty also modify the evolution and the masses of the stars considered.

\subsection{Impact of the mixing length parameter}
As mentioned previously, the mixing-length parameter has a strong effect on the temperature at which the instability strip appears, as it controls the size of the convective zone. Without considering the effect of rotation, we explored the instability strips for two values of the mixing-length parameter: $\alpha_{\mathrm{MLT}} =  2.0,$ and $2.2$.  
\begin{figure}[h]
\centering
\includegraphics[width=\hsize]{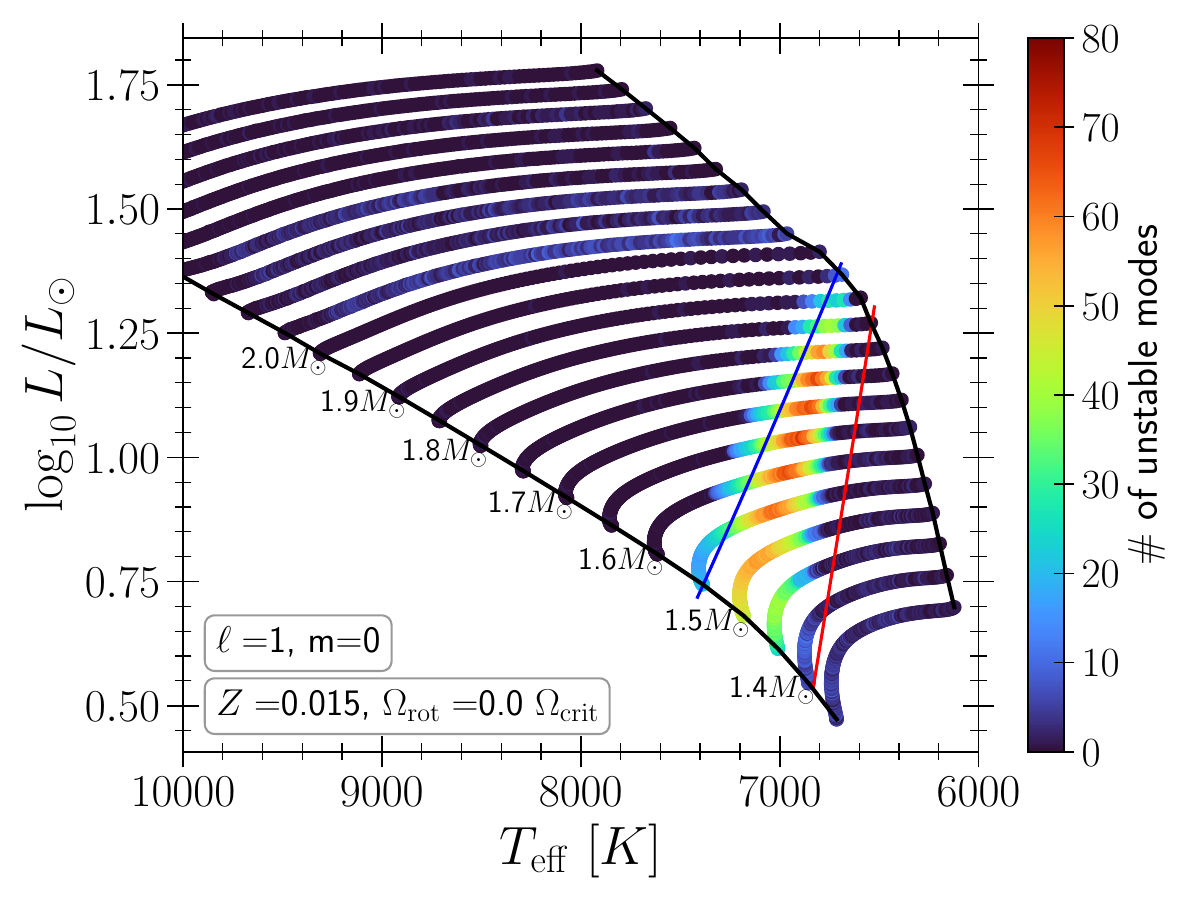}
\caption{HR diagram of the $\gamma$-Doradus instability strip for a grid with $\alpha_{\mathrm{MLT}} = 2.2$. The colour code corresponds to the number of unstable $\ell=1$ $g$ modes for each model in each evolutionary sequence.The blue and red edges are the same as  in \autoref{fig.reference_IS_gammaDor}.}
\label{fig.gammaDor_alphaMLT}
\end{figure}
As shown in \autoref{fig.gammaDor_alphaMLT},  the mixing-length parameter has a strong impact on the instability strip, as already reported by \citet{Dupret2005}. In particular, increasing $\alpha_{\mathrm{MLT}}$ shifts the IS towards hotter models. While the observed population of $\gamma$-Doradus stars might appear to provide constraints on $\alpha_{\mathrm{MLT}}$. However, the treatment of convection–pulsation interactions remains uncertain, and even with TDC models, only an estimate of $\alpha_{\mathrm{MLT}}$ that yields instability strips in approximate agreement with observations for our stellar structure code can be provided. Our purpose here is to illustrate the sensitivity of the instability strip to $\alpha_{\mathrm{MLT}}$, which remains one of the dominant parameters affecting its position in the HR diagram.  %We also notice that the 

\subsection{Impact of the overshooting}
\begin{figure}[h]
\centering
\includegraphics[width=\hsize]{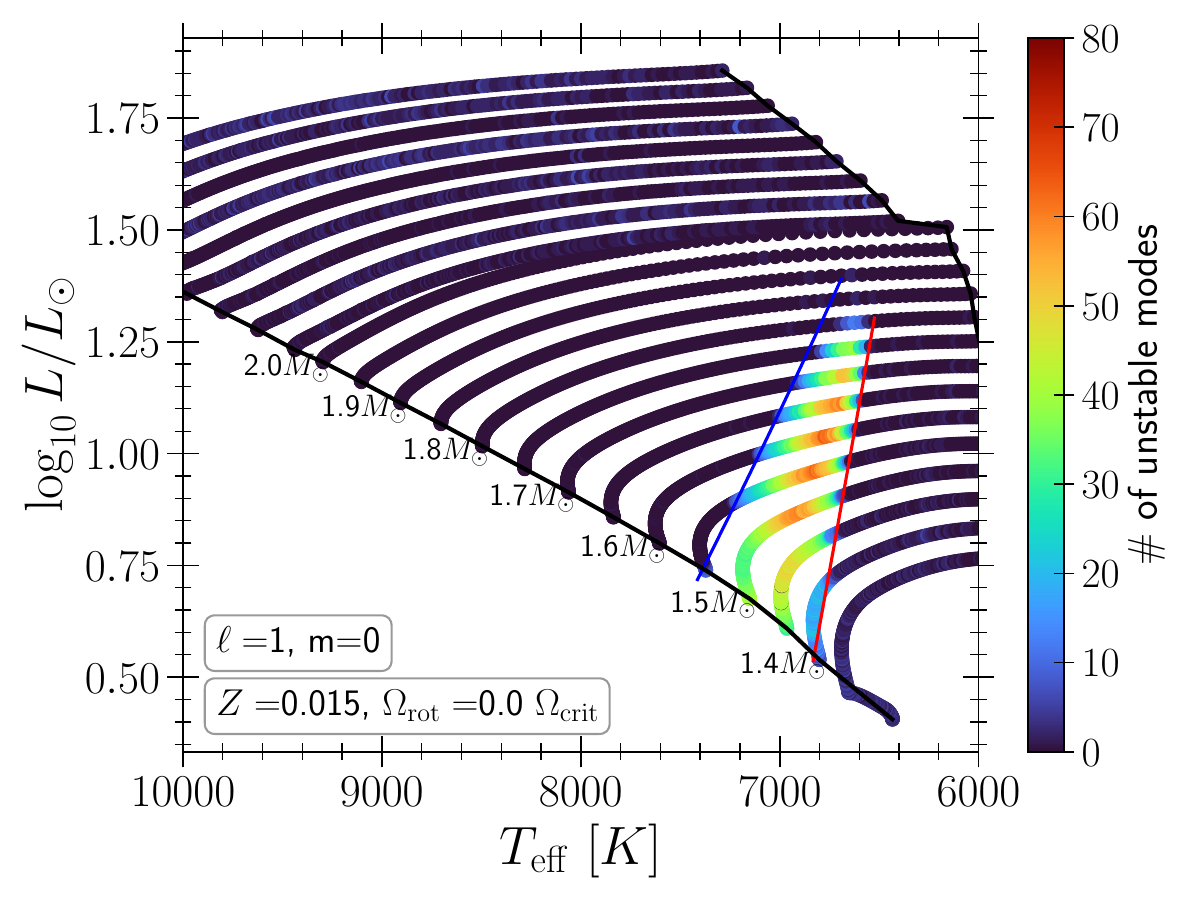}
\caption{HR diagram of the $\gamma$-Doradus instability strip for a grid with $\alpha_{\mathrm{MLT}} = 2.0$ and overshooting parameter $\alpha_{\mathrm{over}}=0.4$. The colour code corresponds to the number of unstable $\ell=1$ $g$ modes for each model in each evolutionary sequence.}
\label{fig.gammaDor_Over}
\end{figure}

One additional parameters can affect the size of convective zones, the overshooting, which controls the extension of the convective core. As overshooting only affects the size of the convective core and thus the stellar evolution, overshooting does not directly impact the effective temperature at which $\gamma$-Doradus stars are found \citep{Grigahcene2006}. To verify this, we extended our grid with models including overshooting, adopting $\alpha_{\mathrm{over}}=0.4$,  where $\alpha_{\mathrm{over}}$ is a step like overshooting parameters extending the convective core of $\alpha_{\mathrm{over}} H_{\mathrm{p}}$ ($H_{\mathrm{p}}$ being the pressure scale height).   In \autoref{fig.gammaDor_Over} we illustrate the $\gamma$-Doradus IS for the case $\alpha_{\mathrm{over}}=0.4$.  
 As expected, we did not find any effect of overshooting on the instability strip, except for the shift in evolutionary tracks caused by the modification of the convective core size. 

\section{Impact of the rotation on the $\gamma-$Doradus IS.} \label{sect_impact_ortation_gammaDor}
\begin{figure}[h]
\centering
\includegraphics[width=\hsize]{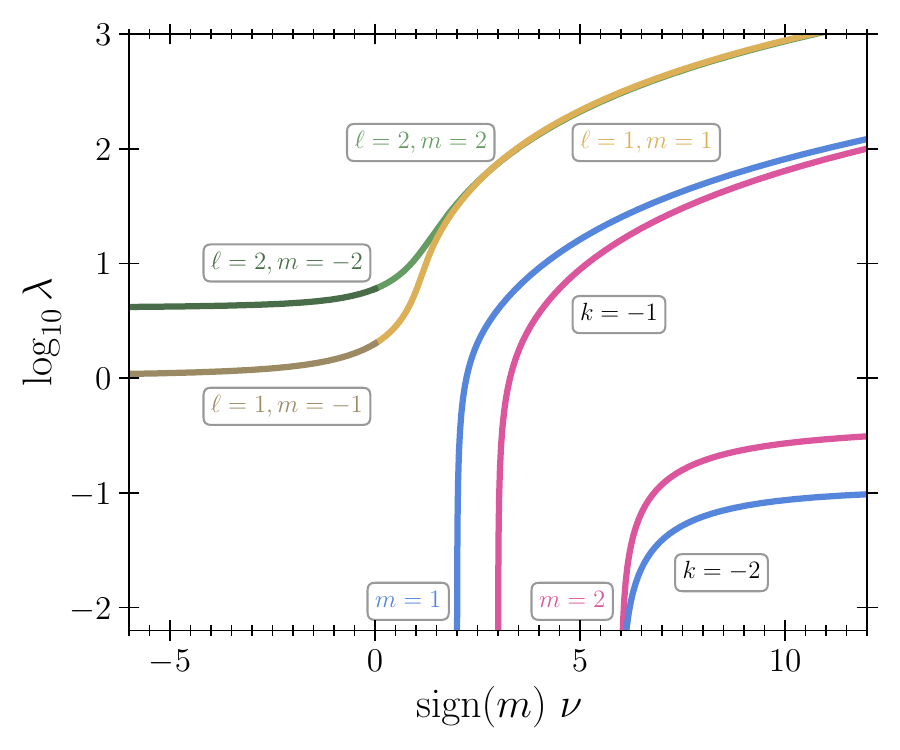}
\caption{Evolution of \(\lambda\) as a function of \(\mathrm{sign}(m)\ \nu\) for modes of different spherical degrees and azimuthal orders. The quantity $\nu$ is the spin parameter defined in \autoref{eq_spin_param}.  Negative values of \(\mathrm{sign}(m)\ \nu\) correspond to prograde modes with \(m<0\), whereas positive values correspond to retrograde modes with \(m>0\). The red and blue curves denote Rossby modes of $m=2$ and $m=1$ respectively, while the green and yellow curves correspond to \(\ell=1\) and \(\ell=2\) gravito-inertial modes, respectively.}
\label{fig.lambda_vs_spin_param}
\end{figure}

\begin{figure}[h]
\centering
\includegraphics[width=\hsize]{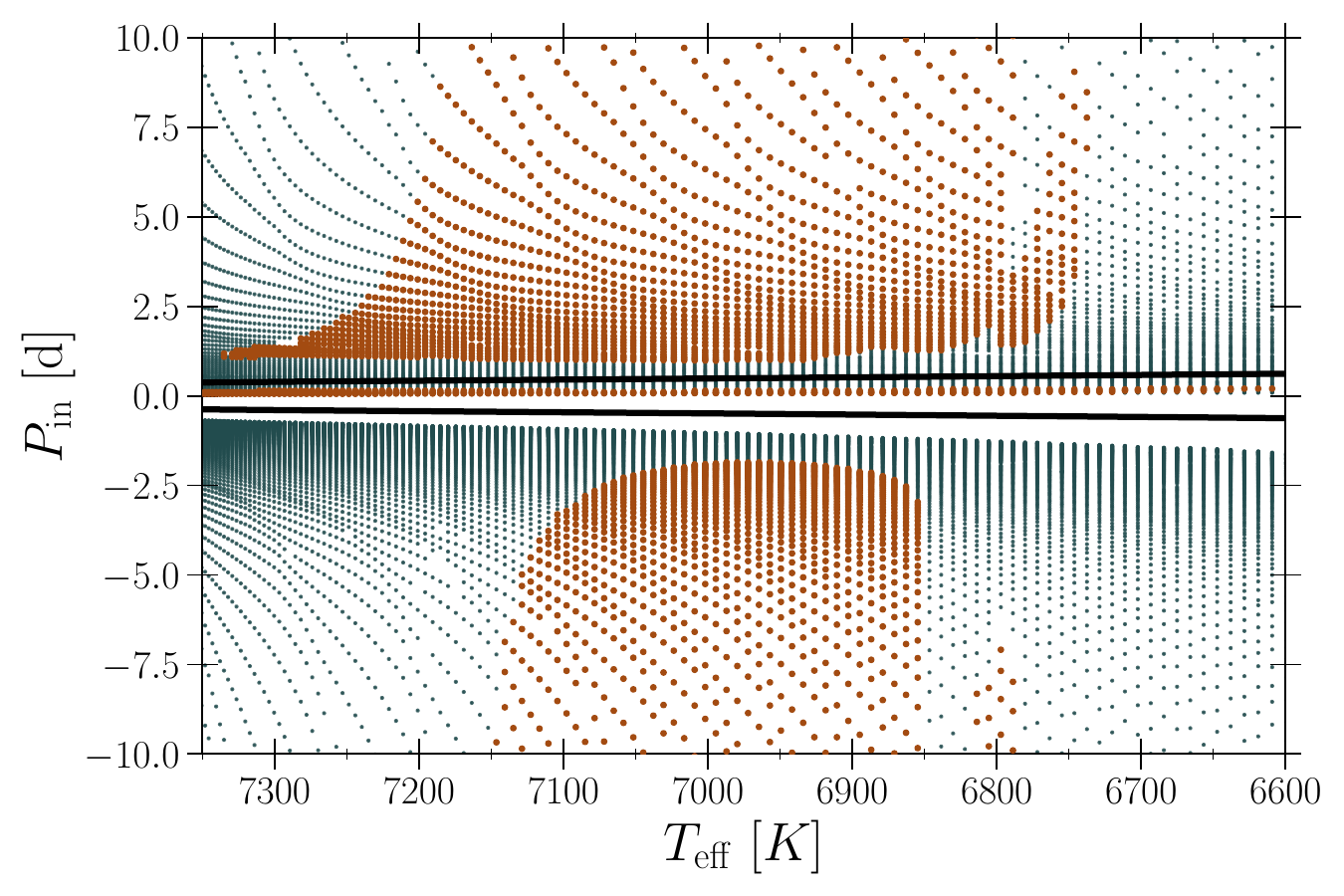}
\caption{Inertial periods of retrograde modes ($\ell=1,m=1$) along the evolution of a $1.55\,M_\odot$ $\gamma$-Doradus star with $\Omega_{\mathrm{rot}}=0.5 \Omega_{\mathrm{crit}}$. Grey dots correspond to stable modes,  orange dots to unstable modes. Modes with negative inertial periods are retrograde modes appearing as prograde in the inertial frame, with observational properties compatible with retrograde modes. The black line corresponds to the evolution of the rotation period in each panel.}
\label{fig.inertial_periods_strip}
\end{figure}

\begin{figure}[h]
\centering
\includegraphics[width=\hsize]{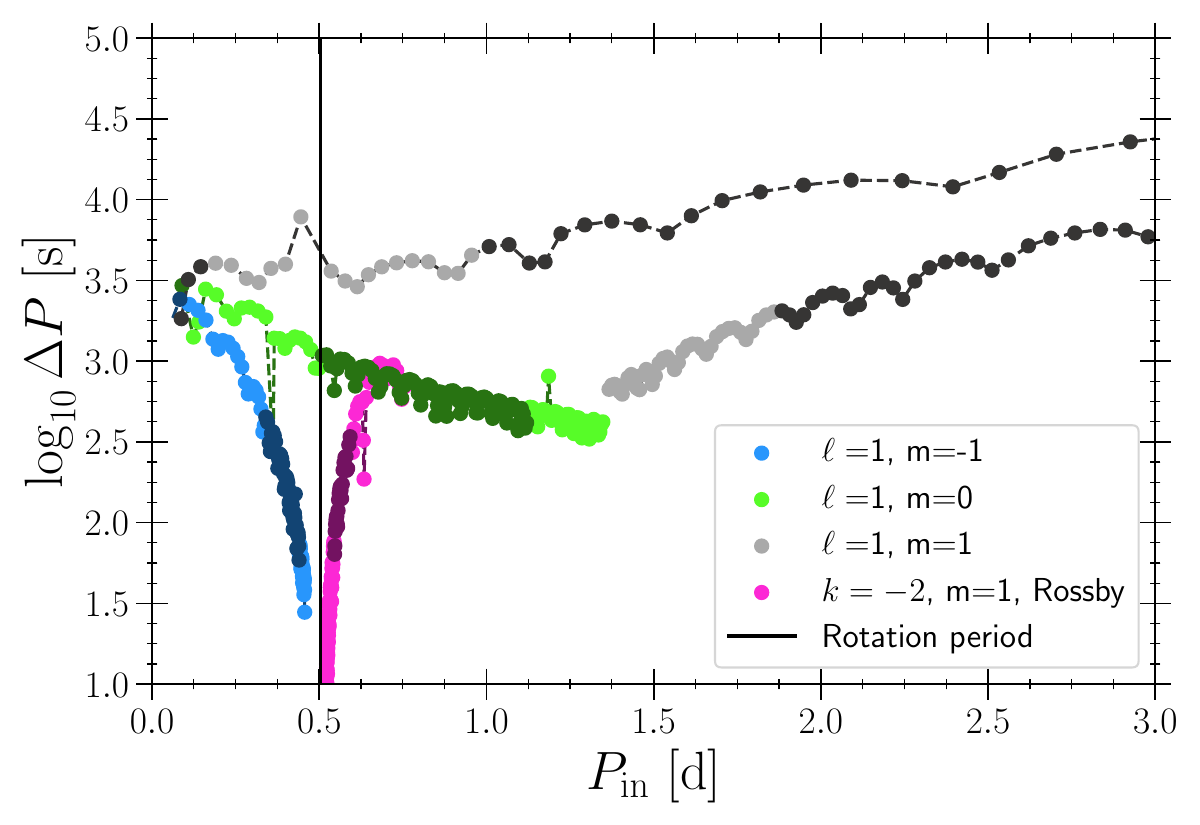}
\caption{Inertial-frame period spacings as a function of the absolute value of the inertial period for a $1.55\,M_\odot$ $\gamma$-Doradus star with $\Omega_{\mathrm{rot}}=0.5 \Omega_{\mathrm{crit}}$ (corresponding to a rotation period of $0.503\ \mathrm{days}$ for this model) and an effective temperature of $6950$ K.  Blue curves: prograde modes. Orange curve: retrograde modes. Green curve: $m=0$ modes. Pink curve: $k=-2, m=1$ Rossby modes. Darker colours correspond to unstable modes.}
\label{fig.deltaP_P_ROT0.5}
\end{figure}

\begin{figure}[h]
\centering
\includegraphics[width=\hsize]{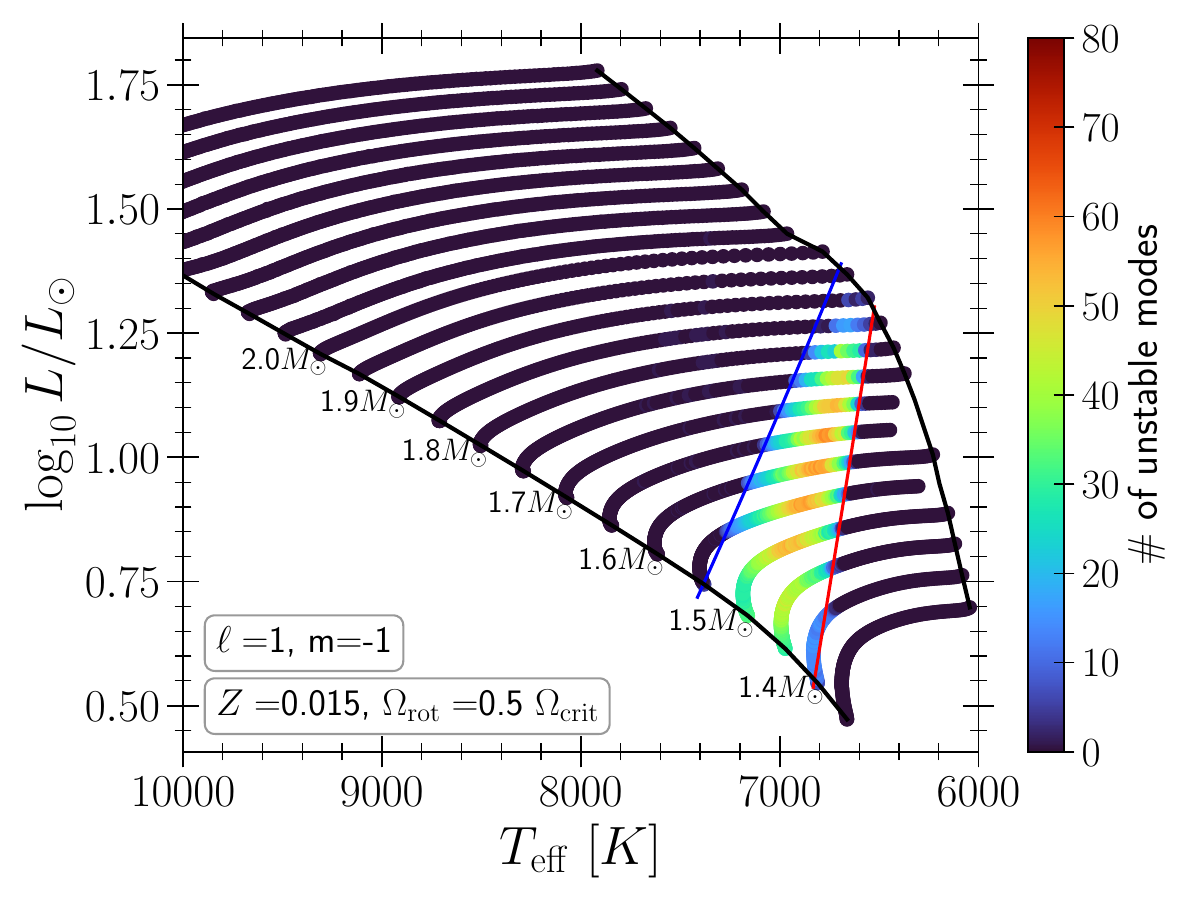}
\caption{HR diagram of the $\gamma$-Doradus instability strip for prograde dipole modes ($\ell=1,m=-1$) with $Z=0.015$ and $\Omega_{\mathrm{rot}}=0.5 \Omega_{\mathrm{crit}}$. The colour code indicates the number of unstable modes.  The blue and red edges are the same as in \autoref{fig.reference_IS_gammaDor}.}
\label{fig.HR_ROT0.5_l1m-1}
\end{figure}

\begin{figure}[h]
\centering
\includegraphics[width=\hsize]{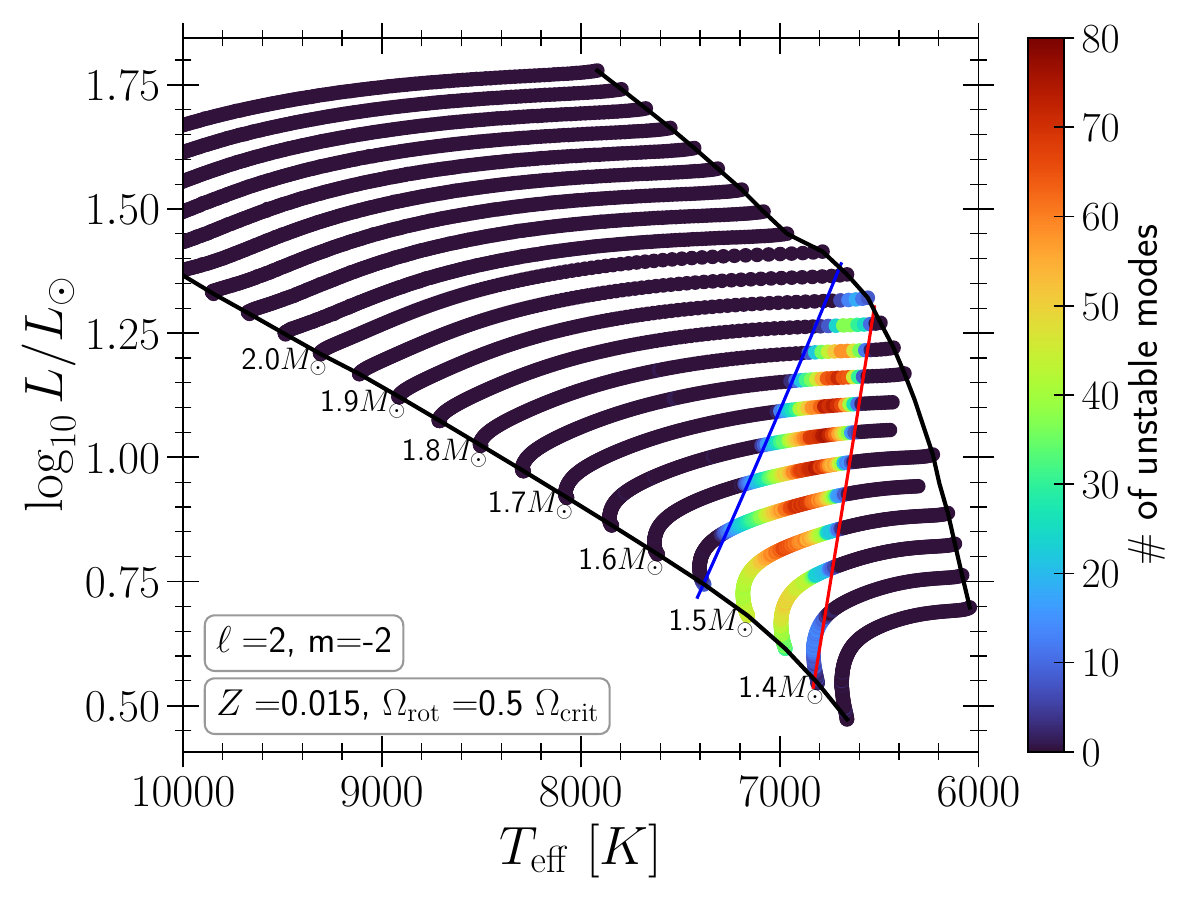}
\caption{HR diagram of the $\gamma$-Doradus instability strip for prograde quadrupole modes ($\ell=2,m=-2$) with $Z=0.015$ and $\Omega_{\mathrm{rot}}=0.5 \Omega_{\mathrm{crit}}$. The colour code indicates the number of unstable modes. The blue and red edges are the same as  in \autoref{fig.reference_IS_gammaDor}.}
\label{fig.HR_ROT0.5_l2m-2}
\end{figure}

\begin{figure}[h]
\centering
\includegraphics[width=\hsize]{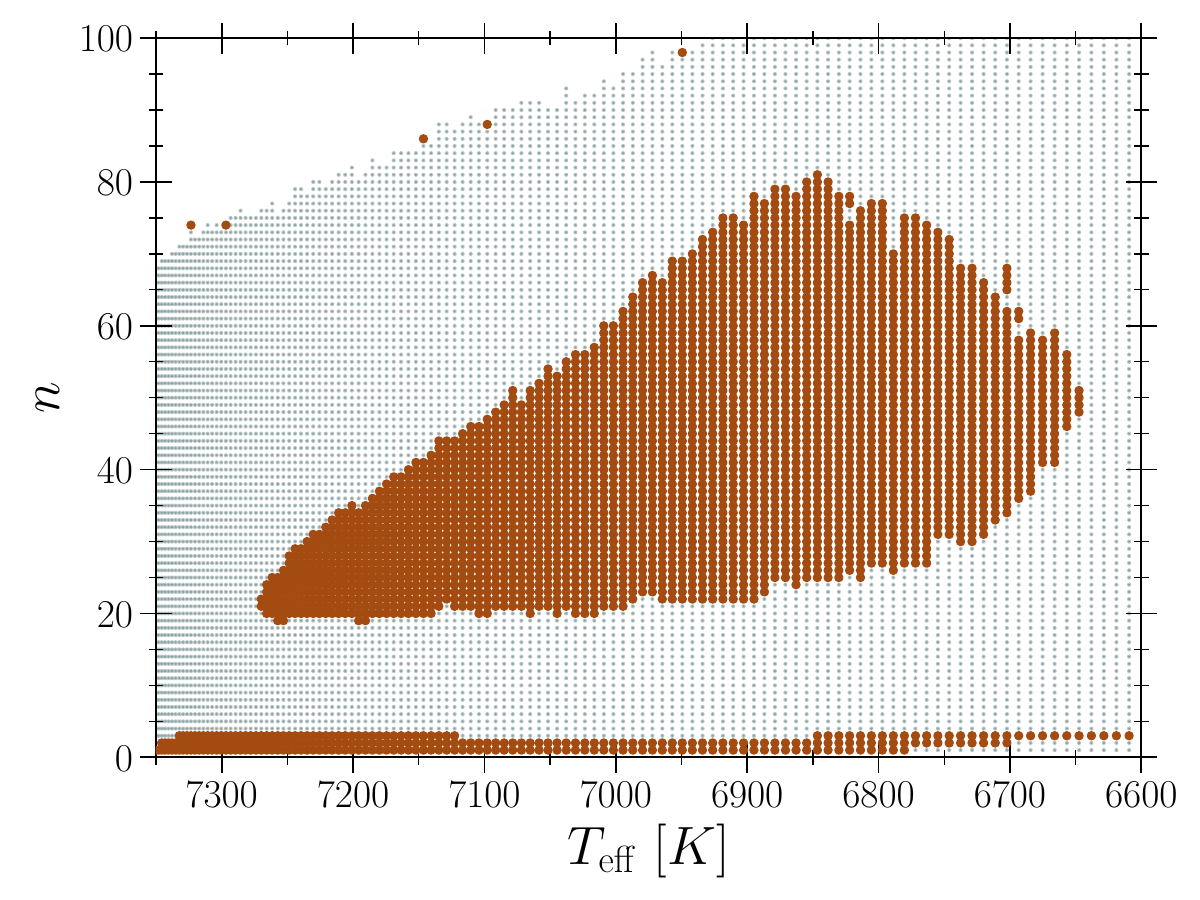}
\caption{Radial orders of stable (black) and unstable (red) $\ell=1,m=-1$ modes along the $1.50\,M_\odot$ evolutionary track with $\Omega_{\mathrm{rot}}=0.5 \Omega_{\mathrm{crit}}$.}
\label{fig._IS_n_Teff_l1m-1ROT0.5}
\end{figure}

\begin{figure}[h]
\centering
\includegraphics[width=\hsize]{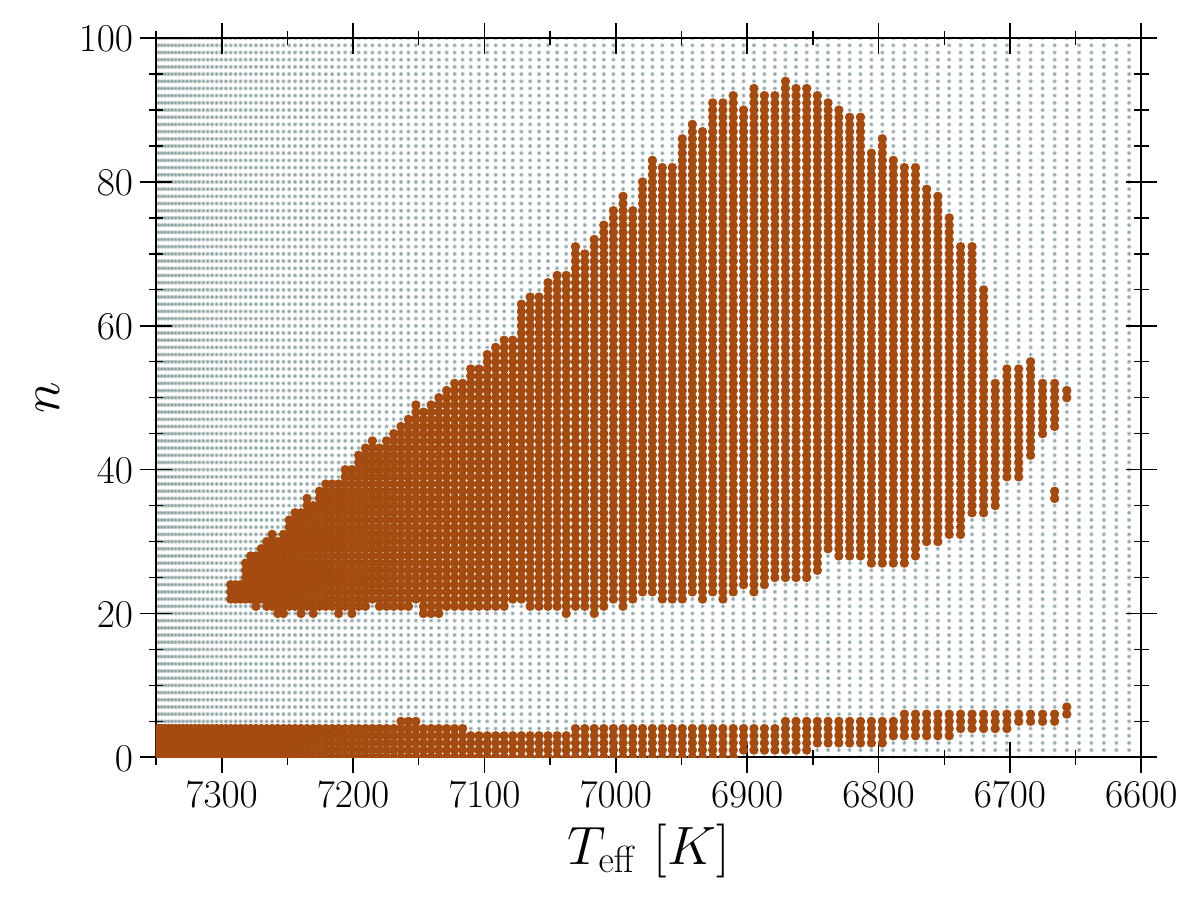}
\caption{Radial orders of stable (black) and unstable (red) $\ell=2,m=-2$ modes along the $1.50\,M_\odot$ evolutionary track with $\Omega_{\mathrm{rot}}=0.5 \Omega_{\mathrm{crit}}$.}
\label{fig._IS_n_Teff_l2m-2ROT0.5}
\end{figure}

As mentioned previously, rotation is expected to impact the instability of oscillation modes in the $\gamma$-Doradus instability strip, in particular by modifying the number of unstable modes and shifting their frequencies depending on the azimuthal order $m$. In addition, rotation is expected to create an instability strip for Rossby modes within the $\gamma$-Doradus domain. Rossby modes are discussed in more detail in \autoref{subsubsect_Rossbygamma}.  In this work we include the impact of the rotation through the traditional approximation for rotation, which is valid for high radial orders gravity modes as in $\gamma$-Doradus stars.
A key consequence of adopting the traditional approximation is that modes are no longer decomposed onto spherical harmonics but onto Hough functions, which depend on rotation, frequency, and azimuthal order \citep{Unno1989}.  Their associated eigenvalue $\lambda$ plays a central role in determining the physical properties and observability of oscillation modes. The larger $\lambda$ is for a given mode, the more confined the mode is to the equatorial regions, reducing its visibility. Thus, we generally expect to favor the observation of modes with low values of $\lambda$.  
The eigenvalue $\lambda$ depends on the spin parameter $\nu$, defined as  
\begin{equation}\label{eq_spin_param}
\nu=\dfrac{2\Omega_{\mathrm{rot}}}{\omega},
\end{equation} 
where $\omega$ is the oscillation frequency in the co-rotating frame and $\Omega_{\mathrm{rot}}$ is the stellar rotation rate.  In rotating stars $\ell$ does not always mean the degree of spherical harmonic, but is used to label a g mode whose amplitude angular dependence becomes spherical harmonic of $\ell$ in the limit of $\Omega_{\mathrm{rot}}$ $\rightarrow$ 0 (where $\lambda$ $\rightarrow$ $\ell(\ell+1)$). The behaviour of $\lambda$ with respect to $\nu$ depends on both the spherical degree $\ell$ and the azimuthal order $m$. This is illustrated in \autoref{fig.lambda_vs_spin_param}.  
As seen in \autoref{fig.lambda_vs_spin_param}, retrograde modes in the corotating frame ($m>0$) display a strong increase in $\lambda$ with increasing spin parameter, while prograde modes in the corotating frame ($m<0$) are much less affected and their $\lambda$ values tend toward a low asymptote. Consequently, in fast-rotating stars, prograde modes are favoured observationally due to their higher visibility, while unstable retrograde modes are more difficult to observe.  On interesting aspect about retrograde modes is that their period in the inertial frame (which corresponds to the frame of the observer),  defined as,
\begin{equation}
P_{\mathrm{in}}=\dfrac{P_{\mathrm{co}}}{1-m \nu/2},
\end{equation}
where $P_{\mathrm{in}}$ is the period in the inertial frame and $P_{\mathrm{co}}$is the period in the corotating frame, can become negative at high spin parameters. In this case,  modes are observed as prograde in the inertial frame, despite behaving as retrograde modes in the corotating frame. This is illustrated in \autoref{fig.inertial_periods_strip}, which shows the inertial periods of retrograde $\ell=1,m=1$ modes in a $1.50\,M_\odot$ $\gamma$-Doradus star rotating at $0.5 \Omega_{\mathrm{crit}}$.  
In this evolutionary sequence, unstable retrograde modes appear observationally as prograde, with periods in the classical $\gamma$-Doradus range. To further illustrate this, \autoref{fig.deltaP_P_ROT0.5} shows the period-spacing patterns for prograde and retrograde modes in fast-rotating $\gamma$-Doradus stars.  
\\~\\As seen in \autoref{fig.deltaP_P_ROT0.5}, classical prograde modes occupy the low-period domain with rapidly decreasing period spacings, while retrograde modes exhibit increasing spacings at longer periods. This behaviour also applies to retrograde modes that appear as prograde (lower dark orange curve). The observed period spacings of many $\gamma$-Doradus stars \citep{Li2020} are well explained by the prograde modes, while retrograde modes are rarely detected. A systematic search for increasing-spacing patterns at longer periods could help identify retrograde modes and thus constrain the internal rotation of fast-rotating $\gamma$-Doradus stars.  Despite this, it should be kept in mind that the observational visibility of retrograde \(g\)-modes is expected to be low due to large  \(\lambda\) values, as shown in \autoref{fig.lambda_vs_spin_param}. The $m=0$ modes form decreasing spacing patterns, between the $m=-1$ and $m=+1$ modes, which is broadly consistent with the few observed cases \citep{Li2020}.  
\\~\\To investigate the impact of rotation more generally, we extended the model grid presented in \autoref{sect_impact_gammaDor} to include rotation under the traditional approximation, assuming solid-body rotation. We considered initial rotation rates from $\Omega=0$ to $0.5\,\Omega_{\mathrm{crit}}$, in steps of $0.1\,\Omega_{\mathrm{crit}}$. Here we focus on prograde modes, which dominate the observations of fast rotators. The full results, including all azimuthal orders, are made publicly available.  
In \autoref{fig.HR_ROT0.5_l1m-1} and \autoref{fig.HR_ROT0.5_l2m-2}, we show the instability strips for prograde $\ell=1$ and $\ell=2$ modes respectively at $\Omega_{\mathrm{rot}}=0.5 \Omega_{\mathrm{crit}}$, for comparison with the reference grid (\autoref{fig.reference_IS_gammaDor}).  
As seen in \autoref{fig.HR_ROT0.5_l1m-1}, the overall position and width of the IS are not strongly affected by rotation, and the number of unstable modes is comparable to the non-rotating case. In contrast, the quadrupole ($\ell=2,m=-2$) modes exhibit more excited modes, as expected from their higher mode density.  
An interesting impact of rotation is on the radial orders of the unstable modes. In \autoref{fig._IS_n_Teff_l1m-1ROT0.5} and \autoref{fig._IS_n_Teff_l2m-2ROT0.5}, we illustrate the radial orders of excited modes as a function of effective temperature for $\ell=1,m=-1$ and $\ell=2,m=-2$ modes, respectively.  
By comparing \autoref{fig._IS_n_Teff_l1m-1ROT0.5} with the reference non-rotating case (\autoref{fig._IS_n_Teff_reference}), we see that rotation shifts the excited modes to lower radial orders for $\ell=1,m=-1$. For quadrupole prograde modes, the IS extends over $n \sim 17$–$84$, broader than for the dipole modes. Observationally, this agrees with \citet{Li2020}, who found a broader radial-order distribution and higher median values for $\ell=2$ modes compared to $\ell=1$ modes.  Overall, the blue edge of the IS tends to be characterised by the excitation of relatively low-order modes ($n\sim15$–40), while the red edge favours higher-order modes ($n\sim30$–60). Near the centre of the IS, modes with $n\sim15$–85 are excited.

\subsection{Rossby modes}\label{subsubsect_Rossbygamma}
\begin{figure}[h]
\centering
\includegraphics[width=\hsize]{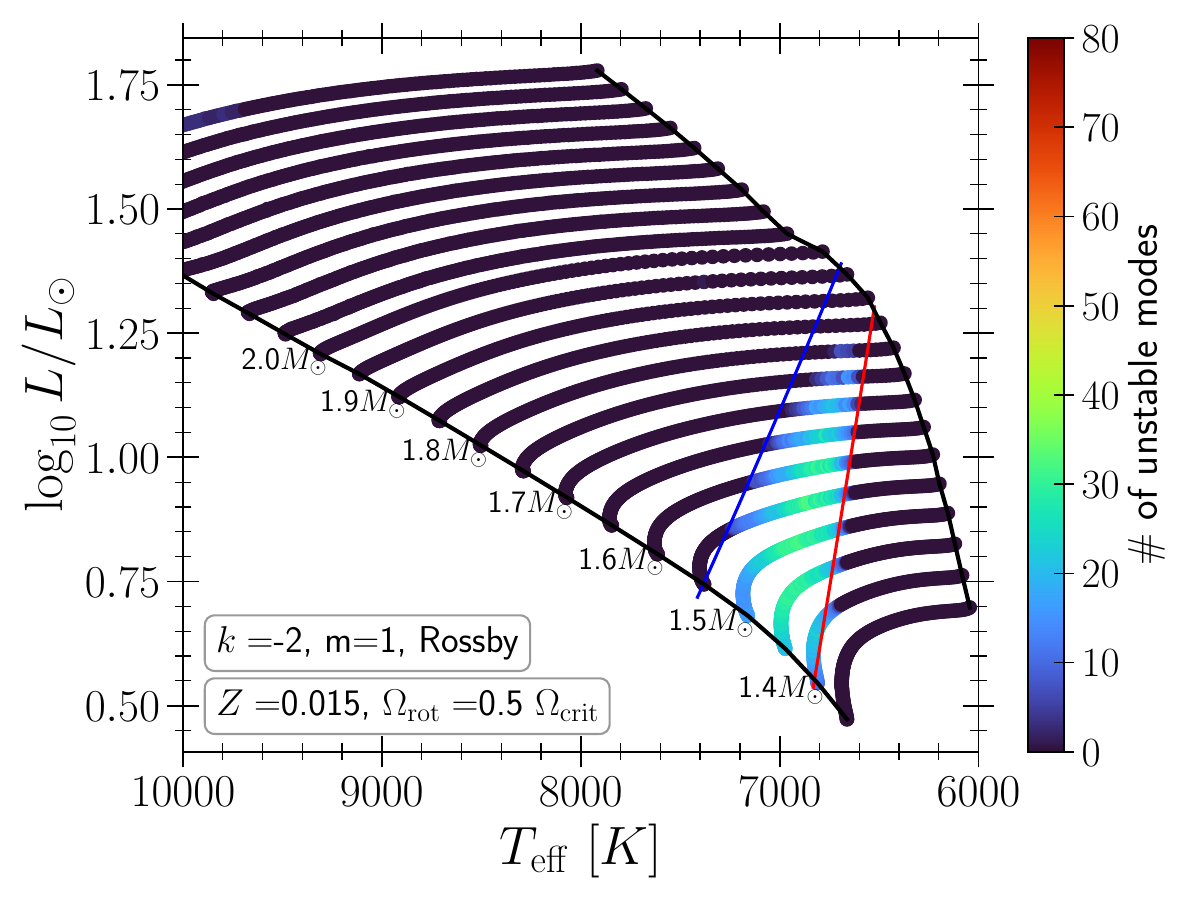}
\caption{HR diagram of the $\gamma$-Doradus instability strip for Rossby modes with $k=-2,m=1$ at $Z=0.015$ and $\Omega_{\mathrm{rot}}=0.5 \,\Omega_{\mathrm{crit}}$. The colour code indicates the number of unstable modes. The blue and red edges are the same as  in \autoref{fig.reference_IS_gammaDor}.}
\label{fig.HR_ROTROS0.2_k-2m1}
\end{figure}

\begin{figure}[h]
\centering
\includegraphics[width=\hsize]{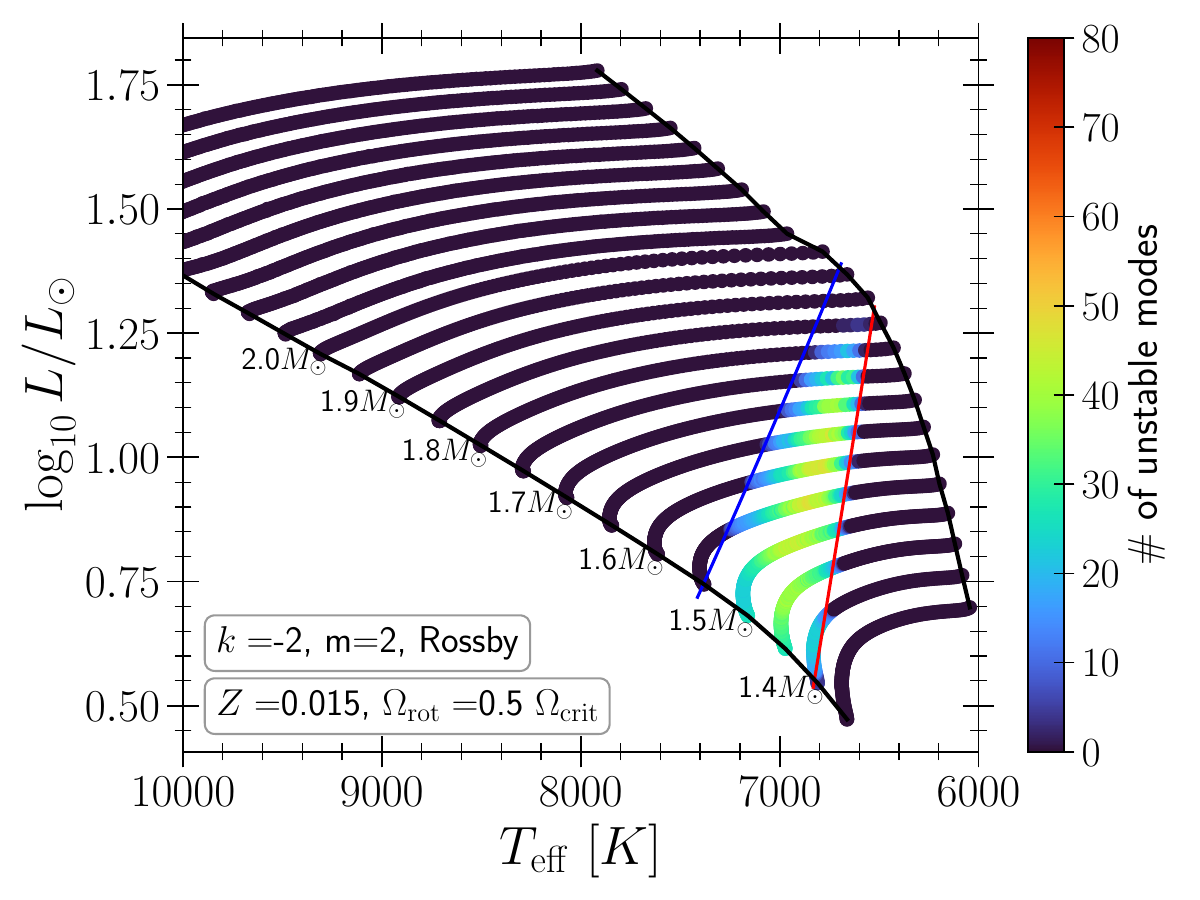}
\caption{HR diagram of the $\gamma$-Doradus instability strip for Rossby modes with $k=-2,m=2$ at $Z=0.015$ and $\Omega_{\mathrm{rot}}=0.5 \,\Omega_{\mathrm{crit}}$. The colour code indicates the number of unstable modes. The blue and red edges are the same as  in \autoref{fig.reference_IS_gammaDor}.}
\label{fig.HR_ROTROS0.2_k-2m2}
\end{figure}

\begin{figure}[h]
\centering
\includegraphics[width=\hsize]{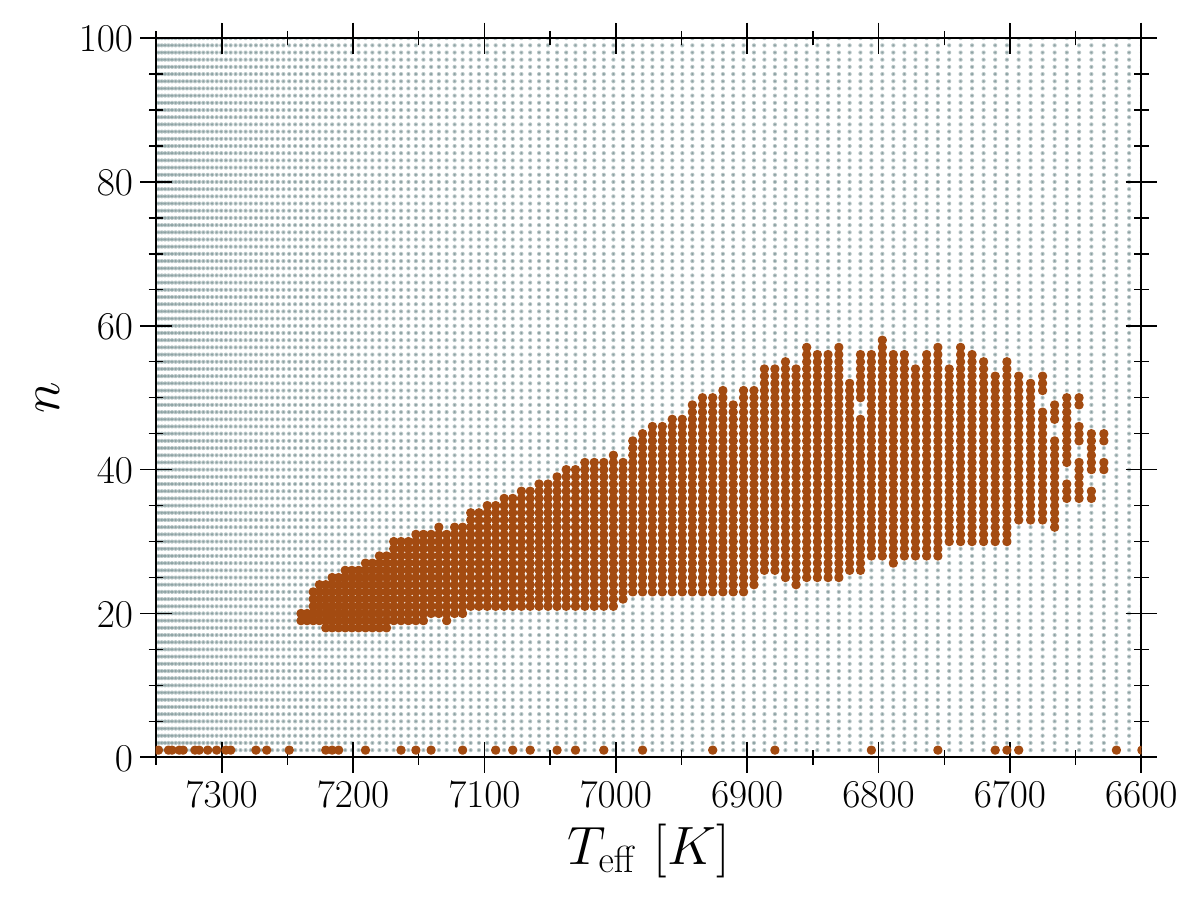}
\caption{Radial orders of stable (black) and unstable (red) Rossby modes with $k=-2, m=1$ along the $1.55\,M_\odot$ reference track with $\Omega_{\mathrm{rot}}=0.5 \,\Omega_{\mathrm{crit}}$.}
\label{fig._IS_n_Teff_ROTROS0.2_k-2m1}
\end{figure}

By including the effect of rotation on the oscillation modes, an additional class of gravito-inertial modes can appear in $\gamma$-Doradus stars: the Rossby modes. Rossby modes are low-frequency, large-scale gravito-inertial oscillations. Their theoretical instability strips were investigated for SPB stars by \citet{Townsend2005, Lee2006}, but have not previously been explored for $\gamma$-Doradus stars. Observationally, \textit{Kepler} data revealed 83 $\gamma$-Doradus stars displaying Rossby modes \citep{Li2020}.  
\\~\\The identification of Rossby modes requires two parameters, $k$ and $m$. For details on $k$, see \citet{Saio2018}. In \autoref{fig.lambda_vs_spin_param}, the $\lambda$ values of Rossby modes are shown in blue and pink for the lowest $k$ and $m$. Despite the higher spin parameters, the corresponding $\lambda$ values are generally lower than those of retrograde gravito-inertial modes, and for $k=-2$ even lower than prograde modes. Since $\lambda$ is a proxy for observability, $k=-2$ modes are more easily detected than $k=-1$. Similarly, $m=1$ Rossby modes are favoured over higher-$m$ values. Observations support this expectation, as nearly all Rossby modes detected by \citet{Li2020} correspond to $k=-2, m=1$.  
In \autoref{fig.HR_ROTROS0.2_k-2m1} and \autoref{fig.HR_ROTROS0.2_k-2m2}, we illustrate the instability strips for the $k=-2, m=1$ and $k=-2, m=2$ Rossby modes, respectively.  
The theoretical IS for the $k=-2, m=1$ Rossby modes is less extended in luminosity than that of the gravito-inertial modes, and the maximum number of unstable modes is shifted toward lower luminosities. This favours the detection of these modes in the lower-mass region of the $\gamma$-Doradus IS. By contrast, the $k=-2, m=2$ Rossby modes occupy a similar HR-diagram region to the gravito-inertial modes. The main difference lies in the number of excited modes: a significantly larger number of $k=-2, m=2$ modes are excited by the flux-blocking mechanism. However,  only the $k=-2, m=1$ modes have been detected in practice \citep{Li2019}.  From these observations, \citet{Li2019} also constrained the radial-order range of the detected Rossby modes. In \autoref{fig._IS_n_Teff_ROTROS0.2_k-2m1}, we show the theoretically predicted range for the $k=-2, m=1$ modes.  
Near the blue edge of the IS, modes with radial orders $n\sim13$–28 are excited, followed by a sharp transition to the main region where a flat distribution of modes with $n\sim15$–45 is excited. Observations by \citet{Li2019} show a similar behaviour: a uniform distribution between $n=10$–15 and $n=45$–50, followed by a gradual decline at higher $n$. Overall the observed and predicted unstable range for Rossby modes seems consistent.

\section{Discussion and conclusion}\label{sect_discussion_conclusion}
In this article we have presented an updated view of the modelling of the $\gamma$-Doradus instability strip (IS), including the effects of metallicity, convective parameters, and rotation. We explored the $\gamma$-Doradus IS under variations of these parameters using a time-dependent convection (TDC) treatment of the convection–oscillation interaction \citep{Dupret2005,Grigahcene2005}. We found that these parameters have only a minor impact on the position of the IS in the HR diagram except for the MLT parameter. However, we showed that rotation affects the excitation of modes and the theoretical period ranges in which the different types of modes can be found. In particular, for fast rotators, retrograde $\ell=1$ modes may appear as prograde in the observer’s frame, with periods of the order or larger than $1.5$~days and a distinctive ascending pattern in the period spacing, in line with previous theoretical and observational studies of gravito-inertial and Rossby modes in intermediate-mass stars \citep{Bouabid2013,VanReeth2015,Saio2018,Li2019}. By including rotation we were also able to explore the Rossby-mode $\gamma$-Doradus IS, which is located in the same region of the HR diagram as the gravito-inertial $\gamma$-Doradus IS.
\\~\\The theoretical exploration of the $\gamma$-Doradus IS also revealed interesting results concerning the radial orders driven by convective blocking when using the TDC treatment. When comparing the range of typically excited radial orders with the mode properties of the 611 clearly identified $\gamma$-Doradus stars observed by \textit{Kepler} and analysed by \citet{Li2019}, we found broad agreement between theory and observations. For prograde modes of both $\ell=1$ and $\ell=2$, our models reproduce the observed ranges of unstable modes, as well as global tendencies such as differences in the median radial orders depending on the mode type. For Rossby modes, where 83 pulsators have been observed, we also found good agreement between predicted and observed radial orders, consistent with previous theoretical expectations for $r$ modes in upper main-sequence stars \citep{Saio2018}.
\\~\\Concerning the position of the theoretical $\gamma$-Doradus IS compared to the observed IS, we still find an extension towards the blue that is not predicted by our models, confirming earlier reports of hot or “outlier’’ $\gamma$-Doradus stars \citep{Balona2016,Qian2019,Antoci2019,Li2019}. Several phenomena could explain this discrepancy. The most straightforward is that the current TDC treatment is insufficient to accurately predict the blue edge of the $\gamma$-Doradus IS. More advanced descriptions of the convection–oscillation interaction may be required, potentially informed by new multi-dimensional convection formalisms and 3D simulations \citep{Dupret2005,Grigahcene2005,Xiong2016,Lizin2024}. However, developing such models is challenging, as it likely requires 3D convection modelling tailored to thin convective envelopes such as those of $\gamma$-Doradus stars.  Other possible explanations for the extended blue edge include misplacement of stars in the HR diagram,  for example due to unresolved binaries that make stars appear hotter or more luminous,  rotational effects impacting the observed surface properties,  and blending or contamination effects in space-based photometry, which can bias the inferred stellar parameters \citep{Murphy2019}.
\\~\\By inspecting the $\gamma$-Doradus IS in more detail, we found that the blue edge predicted by the TDC treatment is characterised by the excitation of relatively low radial-order modes ($n \sim 15$-40 for $\ell=1$), while in the centre of the IS modes with $n \sim 20$-80 are excited, and at the red edge only modes with $n \sim 30$-60 are driven. An analysis of the radial orders of the excited modes of stars observed at the extreme blue edge of the $\gamma$-Doradus IS could be interesting. If such stars excite only low-order modes, they must indeed lie close to the theoretical blue edge, which models currently fail to reproduce. However, if they also excite high-order modes, then these stars could be misplaced in the HR diagram or the TDC treatment is insufficient to accurately predict the range of unstable modes in those regimes. All the computations made for this article and the full grids of $\gamma$-Doradus stars can be found at this address: \url{https://doi.org/10.58119/ULG/F8OXLQ}.

\begin{acknowledgements} 
L.F was supported by the Fonds de la Recherche Scientifique F.R.S-FNRS as a Research Fellow. The authors want to thank the referee for their constructive comments. 
\end{acknowledgements}
\bibliography{Article_prescriptions_GammaDor.bib}
\bibliographystyle{aa}
\newpage
\appendix
\section{Equations for the reference theoretical IS limits }\label{apx_eq_edges_IS}

The theoretical blue edge of the $\gamma-$Doradus IS is defined by the following equation:
\begin{equation}
\log_{10}(L/L_\odot)=0.5660-38.290\left(\log_{10}T_{\mathrm{eff}}-3.834 \right),
\end{equation}
while theoretical red edge is defined as
\begin{equation}
\log_{10}(L/L_\odot)=0.8692-15.083(\log_{10}T_{\mathrm{eff}}-3.860).
\end{equation}

\end{document}